\documentclass[twocolumn]{aastex631}
\newcommand{\hst}{{HST}}

\newcommand{\jwst}{{JWST}}
\newcommand{\lenstool}{\texttt{Lenstool}}

\newcommand{\Lenstool}{{\tt{Lenstool}}}
\newcommand{\wslap}{{\tt{WSLAP+}}}

\newcommand{\purple}{\textcolor{black}}
\newcommand{\referee}{\textcolor{black}}

\newcommand{\Msun}{M$_{\odot}$\,}

\newcommand{\para}{\textbf{Cluster}}
\newcommand{\parb}{\textbf{RA [deg]} }
\newcommand{\parc}{\textbf{Dec [deg]}}
\newcommand{\pard}{\textbf{$\bm\chi^2_{\nu}$}}

\newcommand{\paref}{\textbf{$\nu$,~N$_{\mathbf{Free}}$}}
\newcommand{\parg}{\textbf{sigpos}}
\newcommand{\lstar}{L${_*}$ Gal.}

\newcommand{\Nobserved}{$124$}

\usepackage{hyperref}
\usepackage{amsmath}

\received{March 18, 2025}
\revised{January 8, 2026}
\accepted{January 20, 2026}

\begin{document}

\title{Strong LensIng and Cluster Evolution (SLICE) with JWST: Early Results, Lens Models, and High-redshift Detections}

\author[0000-0002-8261-9098]{Catherine Cerny}
\affiliation{Department of Astronomy, University of Michigan 
1085 South University Avenue 
Ann Arbor, MI 48109, USA}

\author[0000-0003-3266-2001]{Guillaume Mahler}
\affiliation{STAR Institute, Quartier Agora - All\'ee du six Ao\^ut, 19c B-4000 Li\`ege, Belgium}

\author[0000-0002-7559-0864]{Keren Sharon}
\affiliation{Department of Astronomy, University of Michigan 
1085 South University Avenue 
Ann Arbor, MI 48109, USA}

\author[0000-0003-1974-8732]{Mathilde Jauzac}
\affiliation{Centre for Extragalactic Astronomy, Department of Physics, Durham University, South Road, Durham DH1 3LE, UK}
\affiliation{Institute for Computational Cosmology, Durham University, South Road, Durham DH1 3LE, UK}
\affiliation{Astrophysics Research Centre, University of KwaZulu-Natal, Westville Campus, Durban 4041, South Africa}
\affiliation{School of Mathematics, Statistics \& Computer Science, University of KwaZulu-Natal, Westville Campus, Durban 4041, South Africa}

\author[0000-0002-3475-7648]{Gourav Khullar}
\affiliation{Department of Astronomy, University of Washington, Physics-Astronomy Building, Box 351580, Seattle, WA 98195-1700, USA}

\author[0000-0002-0443-6018]{Benjamin Beauchesne}
\affiliation{Institute of Physics, Laboratory of Astrophysics, Ecole Polytechnique Fédérale de Lausanne (EPFL), Observatoire de Sauverny, 1290 Versoix, Switzerland}

\author[0000-0001-9065-3926]{Jose M. Diego}
\affiliation{Instituto de F\'isica de Cantabria (CSIC-UC). Avda. Los Castros s/n. 39005 Santander, Spain}

\author[0000-0002-7633-2883]{David J. Lagattuta}
\affiliation{Centre for Extragalactic Astronomy, Department of Physics, Durham University, South Road, Durham DH1 3LE, UK}
\affiliation{Institute for Computational Cosmology, Durham University, South Road, Durham DH1 3LE, UK}

\author{Marceau Limousin}
\affiliation{Aix Marseille Univ, CNRS, CNES, LAM, Marseille, France }

\author[0000-0001-6804-0621]{Nency R. Patel}
\affiliation{Centre for Extragalactic Astronomy, Department of Physics, Durham University, South Road, Durham DH1 3LE, UK}
\affiliation{Institute for Computational Cosmology, Durham University, South Road, Durham DH1 3LE, UK}

\author[0000-0001-5492-1049]{Johan Richard}
\affiliation{Univ Lyon, Univ Lyon1, ENS de Lyon, CNRS, Centre de Recherche Astrophysique de Lyon UMR5574, 69230 Saint-Genis-Laval, France}

\author[0009-0008-4494-1642]{Carla Cornil-Baïotto}
\affiliation{Instituto de Física y Astronomía, Universidad de Valparaíso, Gran Bretaña 1111, Playa
Ancha, Valparaíso 2360102, Chile}

\author[0000-0003-1370-5010]{Michael D. Gladders}
\affiliation{ Kavli Institute for Cosmological Physics, University of Chicago, 5640 South Ellis Avenue, Chicago, IL 60637, USA}
\affiliation{Department of Astronomy and Astrophysics, University of Chicago, 5640 South Ellis Avenue, Chicago, IL 60637, USA}

\author[0000-0002-7186-7889]{Stephane V. Werner}
\affiliation{Centre for Extragalactic Astronomy, Department of Physics, Durham University, South Road, Durham DH1 3LE, UK}
\affiliation{Institute for Computational Cosmology, Durham University, South Road, Durham DH1 3LE, UK}

\author[0000-0001-5354-4229]{Jessica E. Doppel}
\affiliation{Centre for Extragalactic Astronomy, Department of Physics, Durham University, South Road, Durham DH1 3LE, UK}
\affiliation{Institute for Computational Cosmology, Durham University, South Road, Durham DH1 3LE, UK}

\author[0000-0003-4175-571X]{Benjamin Floyd}
\affiliation{Institute of Cosmology and Gravitation, University of Portsmouth, Burnaby Road, Portsmouth PO1 3FX, UK}

\author[0000-0002-0933-8601]{Anthony H. Gonzalez}
\affiliation{Department of Astronomy, 211 Bryant Space Science Center, University of Florida, Gainesville, FL USA 32611-2055}

\author[0000-0002-6085-3780]{Richard J. Massey}
\affiliation{Centre for Extragalactic Astronomy, Department of Physics, Durham University, South Road, Durham DH1 3LE, UK}
\affiliation{Institute for Computational Cosmology, Durham University, South Road, Durham DH1 3LE, UK}

\author[0000-0001-7847-0393]{Mireia Montes}
\affiliation{Institute of Space Sciences (ICE, CSIC), Campus UAB, Carrer de Can Magrans, s/n, 08193 Barcelona, Spain. }

\author[0000-0003-1074-4807]{Matthew B. Bayliss}
\affiliation{Department of Physics, University of Cincinnati, Cincinnati, OH 45221, USA}

\author[0000-0001-7665-5079]{Lindsey E. Bleem}
\affiliation{High-Energy Physics Division, Argonne National Labora-
tory, 9700 South Cass Avenue., Lemont, IL, 60439, USA}
\affiliation{Kavli Institute for Cosmological Physics, University of
Chicago, 5640 South Ellis Avenue, Chicago, IL, 60637, USA}

\author[0000-0003-1398-5542]{Rebecca E. A. Canning}
\affiliation{Institute of Cosmology and Gravitation, University of Portsmouth, Burnaby Road, Portsmouth PO1 3FX, UK}

\author[0000-0002-3398-6916]{Alastair C. Edge}
\affiliation{Centre for Extragalactic Astronomy, Department of Physics, Durham University, South Road, Durham DH1 3LE, UK}
\affiliation{Institute for Computational Cosmology, Durham University, South Road, Durham DH1 3LE, UK}

\author[0000-0001-5226-8349]{Michael McDonald}
\affiliation{Massachusetts Institute of Technology, 77 Massachusetts Ave, Cambridge, MA 02139, USA}

\author[0000-0002-5554-8896]{Priyamvada Natarajan}
\affiliation{Department of Astronomy, Yale University, New Haven, CT 06511, USA}
\affiliation{Department of Physics, Yale University, New Haven, CT 06511, USA}

\author[0000-0002-2718-9996]{Antony A. Stark}
\affiliation{Center for Astrophysics | Harvard \& Smithsonian, 60 Garden Street, Cambridge, MA 02138, USA}

\author[0009-0004-7337-7674]{Raven Gassis}
\affiliation{Department of Physics, University of Cincinnati, Cincinnati, OH 45221, USA}

%%%%%%%%%%%%%%%%%%%%%%%%%%%%%%%%%%%%%%%%%%%

\begin{abstract}

We leverage \jwst’s superb resolution to derive strong lensing mass maps of 14 clusters, spanning a redshift range of $z\sim0.25 - 1.06$ and a mass range of $M_{500}\sim2-12 \times 10^{14}M_\odot$, from the Strong LensIng and Cluster Evolution (SLICE) \jwst\ program. These clusters represent a small subsample of the first clusters observed in the SLICE program that are chosen based on the detection of new multiple image constraints in the SLICE-\jwst\ NIRCam/F150W2 and F322W2 imaging. These constraints include new lensed dusty galaxies and new substructures in previously identified lensed background galaxies. Four clusters have never been modeled before. For the remaining 10 clusters, we present updated models based on \jwst\ and \hst\ imaging and, where available, ground-based spectroscopy. We model the global mass profile for each cluster and report the mass enclosed within 200 and 500\,kpc. We report the number of new \referee{lensed source galaxies} identified in the \jwst\ imaging, which in one cluster is as high as 19 new \referee{lensed galaxies}.  The addition of new \referee{lensed source galaxies}  and constraints from substructure clumps improves the ability of strong lensing models to accurately reproduce the interior mass distribution of each cluster. We also report the discovery of a candidate transient in a lensed image of the galaxy cluster SPT-CL J0516-5755. All lens models and their associated products are available for download at the
\href{https://data.lam.fr/sl-cluster-atlas/}{Strong Lensing Cluster Atlas Data Base}, which is hosted at Laboratoire d’Astrophysique de Marseille.

\end{abstract}

%%%%%%%%%%%%%%%%%%%%%%%%%%%%%%%%%%%%%%%%%%%


%%%%%%%%%%%%%%%%%%%%%%%%%%%%%%%%%%%%%%%%%%%

\section{Introduction} \label{sec:intro}

Clusters of galaxies are home to the largest gravitationally-bound concentrations of mass in the Universe.
Their deep potentials make them well-suited to exploring the co-evolution and the assembly history of their constituents: dark matter (DM), intracluster plasma (or intracluster gas), galaxies, and free-floating stars (also known as intracluster light).

The exact nature of dark matter, the elusive but vital component of the Universe's total mass budget, is not yet understood. As it has yet to be directly detected, the uncertainty surrounding its nature has left space for several models to develop beyond the current cosmological concordance model of the Universe, which considers cold DM, $\Lambda$-CDM \citep[]{weinberg2013,abott2018,planck2018,perivolaropoulos2022}. 
Galaxy clusters are excellent laboratories in which to test these theories, and to confront observations with model predictions for DM properties that are imprinted on the DM distribution and its spatial correlation with baryonic mass. 
Doing so requires studying the mass distributions in galaxy clusters across cosmic time, which in turn requires high resolution mapping of the mass of galaxy clusters at different redshifts, mass ranges, and densities. 

As DM and observable matter are inextricably linked, the evolution of the distribution of luminous matter across cosmic time can be used as an indirect probe of the evolution of structures and substructures of DM (\citealt{natarajan17}). Adding information from gravitational lensing, and strong lensing (SL) in particular, which relies on observations of background galaxies behind the cluster of interest, provides a more direct probe of the innermost parts of clusters, as it measures the total (dark and baryonic) mass distribution at the smallest spatial scales (e.g. a few kiloparsecs in the cluster plane) in cluster cores \citep{natarajan2024}.

The high resolution and sensitivity of space-based imaging with the Hubble Space Telescope (\hst) revolutionized SL science \citep{kneibnatarajan}. A prime example is the Hubble Frontier Fields program \citep{lotz2017}, which, through investment in hundreds of \hst\ orbits, ground- and space-based spectroscopy, and the entire SL community effort, brought cluster SL accuracy and precision to a new level \citep{johnson2014,richard2014,sharon2015,jauzac2015,diego2016, grillo2016, priewe2017,meneghetti2017,remolina2018, Caminha2019, bergamini2023a, acebron2024}. Other cluster surveys with \hst, as well as targeted observations of single clusters, have demonstrated the power of SL to probe structures at different mass ranges, from the most massive clusters down to groups and even individual galaxies, and to use SL clusters as calibrated cosmic telescopes to magnify and study the Universe behind them \citep{zheng2012,coe2013,zitrin2014,rigby2017,salmon2018,sharon2022}.

SL science has taken yet another step forward in the era of \jwst, owing to \jwst's sensitivity, wavelength coverage, and resolution at long wavelengths, which have enabled the discovery of new multiple images with remarkable precision \citep[e.g.,][]{furtak22, caminha2023, mahler23, bergamini2023c}. The increased number of constraints  resulted in high-resolution mass estimates for galaxy clusters \citep[]{furtak2024}. The results from the first years of SL-enhanced \jwst\ observations remarkably showcase its utility in probing the mass distribution in foreground structures \citep[SMACS+more]{pascale2022,furtak2023,rhitar2024}, discovering and studying galaxies at the highest redshifts \citep[]{atek2023,adamo2024,bradac2024,mowla2024}, and probing star forming regions and stellar clumps at cosmic noon at spatial scales smaller than 100 pc \citep[]{vanzella2022, Claeyssens23, forbes2023,meena2023,emil2024, vanzella2024, claeyssens2025, messa2025, rigby2025}.

In this paper, we introduce the Strong LensIng and Cluster Evolution (SLICE) program (Mahler et al. 2026, in preparation), a large Cycle-3 \jwst~SURVEY program (\#PID: 5594, PI: Mahler) designed to track the evolution of both the luminous and dark components of galaxy clusters over 8\,Gyr of cosmic time, from $z\sim0.2$ to $z=1.9$. The program is obtaining NIRCam imaging of targets drawn from a parent list of 182 clusters, selected primarily for their mass as determined from Sunyaev-Zel'dovich (SZ) effect and X-ray mass proxies, and has at the time of publication observed a total of \Nobserved\ clusters.

Through observations of clusters selected by mass to be on the same evolutionary track, SLICE aims to study: (1) the buildup of the stellar content in the brightest cluster galaxies (BCGs) and the intracluster light \purple{(ICL)}; and (2) the distribution of DM in clusters. The unique sensitivity and resolution of \jwst\ in the infrared reveals old stellar populations, resolving numerous globular star clusters embedded in the ICL (e.g., \citealt{diego2023,martis2024}), in addition to tidal fronts and other structures in the ICL \citep[]{montes2022}. The mapping of DM \purple{will} be measured primarily via SL modeling, but can also be inferred in non-SL clusters from the distribution of \purple{ICL \citep{montesreview2022} and/or} globular clusters, which are \purple{likely} good tracers of the underlying DM density \citep{lee2010,ko2017,yoo2025}, \purple{as shown observationally with HST \citep{montes2019} and in other studies using JWST imaging \citep{lee2022,diego2024,martis2024, cha2025}.~}
The combination of these two science goals will reveal the co-evolution of luminous and dark matter in the densest nodes of the cosmic web. By design, the program benefits from ancillary imaging data from space (\hst\ for all targets, with X-ray for a subsample) and ground-based spectroscopy for many fields.

We present SL models of 14 clusters from SLICE, a first step in achieving the ultimate science goals of this program.
These clusters were chosen from SL clusters observed in the first few months of SLICE observations and span almost the entire range of redshifts, masses, and Einstein radii of the parent sample. Four clusters have never been modeled in the literature using SL techniques, while the remaining clusters have published lens models based on \hst~imaging and other available data. We thus take the opportunity to revisit and update these models with new constraints or new lensing substructure features that are newly identified in \jwst. Most of the new constraints were not visible in \hst\ imaging, either because they are too red (e.g., dusty or high redshift galaxies), or too faint to be robustly identified at the depth of the archival data. The models developed in this work will be made available to the community at the
\href{https://data.lam.fr/sl-cluster-atlas/}{Strong Lensing Cluster Atlas Data Base}, hosted at Laboratoire d'Astrophysique de Marseille.

The paper is organized as follows. Section~\ref{sec:obs} presents the observations and data reduction of \jwst\ and \hst\ observations used to create the SL mass models, and Chandra X-ray imaging of some of the clusters, used for interpretation of the results. Section~\ref{sec:massmodeling} presents the lens models for each cluster, and Section~\ref{sec:results} details the findings for the number counts of the lensing features examined in this paper. In Section~\ref{sec:disc}, we discuss how these models will be used in future SLICE analyses and introduce upcoming work that will be completed once the full SLICE sample has been observed.

We assume a standard $\Lambda$CDM cosmology with $\Omega_M~=~0.3$, $\Omega_{\Lambda}~=~0.7$, and $H_0$~=~70 km s$^{-1}$ Mpc$^{-1}$. All magnitudes are measured in the AB system unless stated otherwise.

%%%%%%%%%%%%%%%%%%%%%%%%%%%%%%%%%%%%%%%%%%%

\begin{deluxetable*}{lrrccccc}
\centering
\tablewidth{0pt}
\tablecaption{SLICE \jwst\ Observations\label{tab.jwst}} 
\tablehead{
    \colhead{Name} & \colhead{$\alpha$ (J2000)} & \colhead{$\delta$ (J2000)} & $z$ &  \colhead{$M_{500}~(10^{14}M_\odot)$} & \nocolhead{Reference} & \colhead{Obs. number} & \colhead{Obs. date} 
}
\startdata
\object[ACO 68]{Abell\,68}   & 9.267227487  &  9.153556343  & 0.2546 & 5.76$^{c}$  & & 16   & 2024-11-26 \\
\object[ACT-CL J0232.2-4421]{RXC\,J0232.2$-$4420}         & 38.0701090   &  -44.3541290  & 0.2836  & 11.30$^{a}$  & & 24  & 2024-10-23  \\
\object[SMACS J2031.8-4036]{SMACS\,J2031.8$-$4036}   & 307.9669300  &  -40.6197000  & 0.3416  & 9.44$^{a}$  & & 37  & 2024-10-14  \\
\object[BAX 006.9575+26.2739]{MACS\,J0027.8$+$2616}    &   6.9662000  &   26.2742000  & 0.365   & 11.70$^{d}$ & & 39   & 2024-08-21  \\
\object[ACT-CL J2211.7-0349]{RXC\,J2211.7$-$0349}     & 332.9307646  &  -3.82922895  & 0.397  & 12.31$^{c}$ & & 52   & 2024-11-18  \\
\object[ACT-CL J0553.4-3342]{MACS\,J0553.4$-$3342}    &  88.3512360  & -33.71225900	& 0.412 & 11.33$^{a}$ & & 55 & 2024-10-30 \\
\object[ACO 3192]{MACS\,J0358.8$-$2955}    &  59.7201370  & -29.92988500  & 0.425  & 8.81$^{a}$  & & 58   & 2024-10-23  \\
\object[ClG J1621+3810]{MACS\,J1621.4$+$3810}    & 245.3500000  &  38.16740000  & 0.4631 & 14.48$^{d}$ & & 71   & 2024-08-19  \\
\object[ACT-CL J0327.4-1326]{RCS2\,032727$-$132623}   &  51.8632708  & -13.43938060  & 0.564  &  5.53$^{b}$    & & 86   & 2024-09-30  \\
\object[ClG J2129-0741]{MACS\,J2129.4$-$0741}    & 322.3544372  &  -7.70892166  & 0.589  & 7.06$^{c}$  & & 89   & 2024-10-03  \\
\object[PSZ1 G091.82+26.11]{PSZ1\,G091.83$+$26.11}     & 277.7835360  &  62.24809379  & 0.822  & 7.43$^{c}$  & & 118  & 2024-09-08  \\
\object[ClG J0152-1358]{RX\,J0152.7$-$1357}        & 28.17764473  & -13.95911996  & 0.8269 & 3.47$^{b}$  & & 119  & 2024-08-26  \\
\object[SPT-CL J0516-5755]{SPT-CL\,J0516$-$5755}        & 79.2397540   & -57.91668800  & 0.9656 & 3.42$^{a}$ & & 139 & 2024-08-29 \\ 
\object[SPT-CL J2011-5228]{SPT-CL\,J2011$-$5228}        & 302.7786200  & -52.47247800  & 1.064  & 3.10$^{a}$  & & 149  & 2024-09-06 \\ 
\enddata
\tablecomments{Summary of \jwst~observations from the SLICE program (PID: 5594; PI: Mahler).  The exposure time for all \jwst\ images in both the NIRCam/F150W2 and NIRCam/F322W2 filters is 1836 s. Column (1): the name of the cluster. Columns (2) and (3): the R.A., $\alpha$, and Decl., $\delta$, are given in degrees. Column (4): the cluster redshift. Column (5): the  $M_{500}$ mass, where the superscript indicates the reference used in our cluster selection, as below. Column (6): the SLICE program visit ID. Column (7): the observation date for the exposure. The superscripts and references for Column (5) are as follows:
$^{a}$ \citealt{Bocquet2019}, $^{b}$\citealt{Hilton2021}, $^{c}$ \citealt{Planck2016}, $^{d}$ \citealt{Repp2018} converting the X-ray flux using Equation (2) from the formula presented in \citealt{Anderson2015}.}
\end{deluxetable*}

\begin{deluxetable*}{lcccccc}[ht]
\tablecaption{Archival \hst\ Observations\label{tab.hst}}
\tablehead{
    \colhead{Name} & \colhead{Band} & \colhead{PID} & Program Type &  \colhead{Exp. time [s]} & \colhead{Obs. date} 
}
\startdata
Abell\,68            & ACS/F814W  & 11591 & GO   & 4880  & 2010-11-14  \\
                    & WFC3/F110W & 11591 & GO   & 2612  & 2010-10-18  \\
                    & WFC3/F160W & 11591 & GO   & 2412  & 2010-10-18  \\
RXC\,J0232.2$-$4420     & ACS/F606W  & 14096 & GO/Treasury$^{1}$   & 2075  & 2016-08-14  \\ 
                    & ACS/F814W  & 14096 & GO/Treasury$^{1}$   & 2246  & 2016-07-05  \\ 
SMACS\,J2031.8$-$4036  & ACS/F606W  & 12166 & SNAP & 1200  & 2010-09-17  \\ 
                    & ACS/F814W  & 12884 & SNAP & 1440  & 2012-11-14  \\
                    & WFC3/F110W & 12166 & SNAP & 706   & 2012-06-21  \\
                    & WFC3/F140W & 12166 & SNAP & 706   & 2012-06-21  \\
MACS\,J0027.8$+$2616   & ACS/F606W  & 12166 & SNAP & 1200  & 2010-10-02  \\
                    & ACS/F775W  & 9770  & GO   & 14370 & 2004-01-17 \\ 
                    & WFC3/F110W  & 12166  & SNAP & 706 & 2010-11-18 \\ 
                    & WFC3/F140W  & 12166  & SNAP   & 706 & 2010-11-18 \\ 
RXC\,J2211.7$-$0349    & ACS/F606W  & 12166 & SNAP & 1200  & 2011-11-19  \\
                    & ACS/F606W  & 14096 & GO/Treasury$^{1}$   & 2101  & 2015-11-25  \\
                    & ACS/F814W  & 14096 & GO/Treasury$^{1}$   & 2124  & 2015-10-16  \\
MACS\,J0553.4$-$3342   & ACS/F435W  & 12362 & GO   & 4452  & 2012-01-05 \\
                    & ACS/F606W  & 12362 & GO   & 2092  & 2012-01-05 \\
                    & ACS/F814W  & 12362 & GO   & 4572  & 2012-01-05 \\
MACS\,J0358.8$-$2955   & ACS/F435W  & 12313 & GO   & 4500  & 2011-02-19  \\ 
                    & ACS/F606W  & 12313 & GO   & 2120  & 2011-02-19  \\ 
                    & ACS/F814W  & 12313 & GO   & 4620  & 2011-02-19  \\ 
MACS\,J1621.4$+$3810   & ACS/F606W  & 12166 & SNAP & 1200  & 2011-07-21  \\
                    & ACS/F814W  & 12166 & SNAP & 1440  & 2011-09-08  \\
                    & WFC3/F110W & 12884 & SNAP & 706   & 2012-10-12  \\
                    & WFC3/F140W & 12884 & SNAP & 706   & 2012-10-12  \\
RCS2\,032727$-$132623  & ACS/F814W  & 12371 & GO   & 1080  & 2011-06-28  \\
                    & WFC3/F125W & 12267 & GO   & 862   & 2011-03-01  \\                    
                   & WFC3/F160W & 12267 & GO   & 862   & 2011-03-01  \\
MACS\,J2129.4$-$0741   & ACS/F606W & 12100 & GO/Treasury$^{2}$ & 3728 & 2011-05-15 \\
                    & ACS/F814W & 10493 & GO & 2168 & 2005-06-18 \\
PSZ1\,G091.83$+$26.11   & ACS/F606W  & 15132 & SNAP & 1200  & 2018-08-27  \\
                    & ACS/F814W  & 14098 & SNAP & 1200  & 2016-07-24  \\
RX\,J0152.7$-$1357      & ACS/F625W  & 9290 & GTO/ACS & 4750 & 2002-11-06  \\
                    & ACS/F160W  & 14096 & GO/Treasury$^{1}$   & 1209  & 2017-01-26 \\
SPT-CL\,J0516$-$5755    & ACS/F606W  & 13412 & SNAP & 2320  & 2014-02-27 \\
SPT-CL\,J2011$-$5228    & UVIS1/F606W & 14630 & GO  & 2690   & 2017-03-18 \\
                    & UVIS2/F606W & 14630 & GO  & 2676   & 2017-03-22 \\
\enddata
\tablecomments{Summary of primary archival \hst~observations that were used in this work. Column (1): the name of the cluster. Column (2): the camera and filter. Column (3): the PID for each observation. Column (4): the program type- General Observer (GO), Snapshot (SNAP), Large Treasury (GO/Treasury), or Guaranteed Time Observations allocated to the ACS team (GTO). Treasury programs are marked with a superscript “1,” referring to RELICS (\citealt{coe2019}), or with a superscript “2,” referring to CLASH (\citealt{postman2012}). Column (5): the exposure time. Column (6): the observation date. The list is not exhaustive: more data are available in the archive for many of the fields. } 
\end{deluxetable*}

\begin{deluxetable}{lcccccc}
\tablecaption{Archival \textit{Chandra} Observations\label{tab.xray} }
\tablehead{
    \colhead{Cluster} & \colhead{Obs. ID} & \colhead{Exp. time [ks]} & Obs. date  
}
\startdata
RXCJ0232.2$-$4420  &  4993 &  23.40  &  2004-06-08 \\
MACS\,J0027.8$+$2616  &  3249    &   9.98  & 2002-06-26  \\
 &  14012 & 21.69 & 2012-09-24 \\
MACS\,J0553.4$-$3342 &  5813 &  9.94  &  2005-01-08 \\
 &   12244 & 74.06 & 2011-06-23 \\
MACS\,J0358.8$-$2955   & 11719 & 9.65 & 2009-19-18 \\
 & 12300 & 29.66 & 2010-11-26 \\
 & 13194 & 19.97 & 2010-11-28 \\
MACS\,J1621.4$+$3810    &   3254 & 9.85 & 2002-10-18 \\
 &   3594 & 19.73 & 2003-08-22 \\
 &   6109 & 37.55 & 2004-12-11 \\
 &   6172 & 29.75 & 2004-12-25 \\
 &   9379 & 29.91 & 2008-10-17 \\
 &  10785 & 29.75 & 2008-10-18 \\
 MACS\,J2129.4$-$0741   &  3199 & 19.86 & 2002-12-23 \\
 & 3595 & 19.87 & 2003-10-18 \\ 
PSZ1\,G091.83$+$26.11   &  18285   &  22.7      &  2016-06-19  \\
RX\,J0152.7-1357   & 914 & 36.48 & 2000-09-08 \\
 & 21703 & 39.56 & 2019-09-25 \\
 & 22856 & 19.28 & 2019-09-26 \\
\enddata
\tablecomments{Summary of archival Chandra X-ray observations, available only for 8 clusters.}
\end{deluxetable}

\section{Observations} \label{sec:obs}
\subsection{\jwst~SLICE Imaging}

The SLICE \jwst~SURVEY mode program (PID: 5594; PI: Mahler) consists of NIRCam imaging using the filters F150W2 and F322W2, which are extrawide filters in the short-wavelength and long-wavelength channels, respectively, and which together cover the wavelength range between 1 to 4 micrometers. The program is designed as a large survey with 182 possible targets, with \Nobserved\ observed at the time of the publication of this paper. The observations are run in the Shallow 4 mode with only Module B active to reduce the volume of the data products that are downloaded from the spacecraft. An IntramoduleX 3 dithering pattern is employed to remove any chip gaps, and an additional 3 point dithering pattern is employed using small-grid-dither to improve the point-spread-function sampling at shorter wavelengths. The observing time is calculated such that a S/N~$\sim$~3 is achieved for a point source at 29 mag, which equates to an equivalent exposure time of 1836\,s in both the F150W2 and the F322W2 bands. This exposure time is consistent across all images in the SLICE program. Both the F150W2 and the F322W2 bands are observed simultaneously. \jwst~observations for the clusters studied in this paper are presented in Table~\ref{tab.jwst}, along with basic information about each cluster, including their redshifts and a measurement for $M_{500}$ that is drawn from various literature references, where $M_{500}$ is defined as the mass within a radius of $R_{500}$ that has an overdensity of 500 times the critical mass density of the Universe at the cluster redshift. 

We follow the procedure described in \cite{Rigby2023} for NIRCam data reduction (as used for the \jwst~Early Release Science program TEMPLATES), which we describe briefly below. We use Level 1b products from MAST, using an STScI-adapted custom Python script. For data reduction, we use scripts modified from the default \jwst~Version (v1.15.1; \citealt{bushouse_2024_12692459}) and the Calibration Reference Data System pipeline mapping (pmap) 1303. We process Level 2A data products to apply a custom de-striping algorithm to correct for 1/f noise and jumps between amplifiers. A complete description of this algorithm will be provided in the forthcoming survey paper (Mahler et al. 2026, in preparation). We then run the de-striped images through the same modified pipeline, to generate Level 3 F150W2 and F322W2 dithered science-ready mosaics. The filters are WCS-matched to Gaia as per the specifications of the default pipeline; \hst~data are aligned to \jwst~imaging within uncertainties. For more details on the various steps and systematics involved in this process, please refer to the complete description provided in \cite{Rigby2023}.

\subsection{\hst~Archival Imaging}
Archival \hst~images for all the clusters presented in this paper were obtained from MAST. For SLICE targets that were previously observed as part of \hst~treasury programs (ReIonization Lensing Cluster Survey -- RELICS, \citealt{coe2019}; Cluster Lensing And Supernova survey with Hubble -- CLASH, \citealt{postman2012}), we used the reduced data that were made available to the community by these projects. For all other targets, we reduced the \hst~imaging data to match the footprint, pixel scale, and WCS solution of the new \jwst~data, using standard procedures as part of {\texttt{Astrodrizzle}}. 
We combined the calibrated sub-exposures taken with each filter ({\texttt{\_flt}} for WFC3-IR, and {\texttt{\_flc}} files for UVIS and the Advanced Camera for Surveys (ACS)) using a Gaussian kernel with a drop size {\texttt{final\_pixfrac}}=$0.8$ and {\texttt{final\_refimage}} set to the NIRCam F150W2 exposure. If needed, images were further aligned to better match the WCS solution of the NIRCam data using {\texttt{tweakreg}} and {\texttt{tweakback}}.  
A complete list of the \hst~images used in this paper is presented in \autoref{tab.hst}. 

\subsection{\textit{Chandra} X-ray Imaging}
We retrieved archival X-ray observations from the \textit{Chandra X-ray Observatory} for a subsample of the clusters. These observations used the Advanced CCD Imaging Spectrometer (ACIS) and are composed of observations obtained with the ACIS-I or ACIS-S configuration. Due to the limited number of counts for specific clusters, we included data obtained in FAINT mode in addition to the more robust VFAINT mode. We reduced the data with the routine implemented in the CIAO software environment \citep[version 4.15]{fruscione2006}. In particular, we used the \textsc{WAVDETECT} source-detection routine \citep{freeman2002} to automatically remove the X-ray point sources and create count maps in the $[0.5,7]$~keV band from the diffuse X-ray emission. We additionally employ the \texttt{csmooth} routine, which is a \textit{Chandra}-focused adaptation of the ASMOOTH routine \citep{csmooth}, to smooth the X-ray maps for the purpose of visual presentation. A complete list of the X-ray imaging used in this paper is given in \autoref{tab.xray}.

%%%%%%%%%%%%%%%%%%%%%%%%%%%%%%%%%%%%%%%%%%%

\section{Strong Lensing Mass Modeling} \label{sec:massmodeling}

In this section, we present SL models that are designed to recover the mass within the core of the observed galaxy clusters ,using the positions of multiply-imaged galaxies as constraints. These lens models are constructed primarily using \lenstool~\citep{jullo07}, a parametric \purple{Markov chain Monte Carlo (MCMC)} modeling software which is described in more detail below. We also present lens models created using the hybrid modeling technique \wslap~\citep{Diego2005,Diego2007} for the galaxy clusters RCS2\,032727$-$132623 (RCS~0327) and PSZ1\,G091.83$+$26.11, which each have a large amount of lensed substructure visible in the \jwst\ imaging that enable the use of free-form methods like \wslap. We discuss the implementation of these two modeling algorithms in the subsections for each cluster (\autoref{sec.RCS0327} and~\autoref{sec.PSZG091}). 

Regardless of the algorithm used, all SL modeling techniques attempt to find a foreground mass distribution that satisfies the lens equation everywhere in the field, and reproduces the observed locations of multiple images of each background source. The multiple images are therefore used as constraints in the lens modeling process. 
Multiple images are identified in the imaging data, usually by eye, owing to their similarity to each other. Since lensing is wavelength-invariant, counter-images should have the same colors (unless they are contaminated by foreground sources or obscured by dust). The morphology of two counter-images should be the same, after accounting for their different distortions, magnifications, and parity. Finally, the parity (e.g., mirroring) should be consistent with the expectations from SL geometry. 

\purple{Counter-images and arc candidates are verified through spectroscopy.} In a few cases, multiple images of the same background source were identified spectroscopically in integral field unit (IFU) observations of the entire lensing field, such as \purple{with the Multi-Unit Spectroscopic Explorer (MUSE) on the Very Large Telescope (VLT)} \citep{bacon2010}. These identifications may or may not have an optical counterpart at the depth of the \jwst/\hst~imaging.
The superior resolution of space-based imaging allows for identification and mapping of structure within multiple images, and for mapping this substructure from one image to another. Substructure is most often in the form of emission clumps, but can also be distinct dust lanes, spiral arms, or other features \citep[e.g.][]{grillo2016, bergamini2023a,lagattuta2023}. Where identified, the substructure can be used to increase the number of constraints.
\autoref{tab.MultipleImages} tabulates the multiple images that were identified in each of the clusters presented in this paper. Where available, we list the spectroscopic, photometric, or model-optimized redshift, and multiple clumps within each image.

\subsection{\lenstool}\label{sec.lenstool}
\lenstool\ is a parametric SL modeling algorithm, which uses MCMC formalism to explore the parameter space, identify the best fit set of parameters, and estimate statistical uncertainties on the parameters and modeling outputs. We refer the reader to \cite{jullo07} for a full description of \lenstool. In short, the algorithm assumes that the foreground lens is comprised of a combination of mass halos, each described by a set of parameters. Some examples of parameterized mass halos include \purple{the Navarro-Frenk-White profile;} isothermal \purple{ profiles;} and, most commonly used for clusters, the \purple{truncated} pseudo-isothermal ellipsoidal mass distribution (dPIE; \citealt{Eliasdottir2007}; also \purple{sometimes} referred to in the literature \purple{simply} as PIEMD). Lensing constraints come from the positions and redshifts of sets of multiple images of the same lensed source. Where spectroscopic redshifts are not available, one can leave the redshift as a free parameter. Photometric redshifts or other information can be used to guide the posterior.
The goodness-of-fit criterion is the image-plane scatter between the observed and predicted images of lensed sources (image-plane minimization), or the source-plane scatter between sets of predicted source locations of the same \referee{sources}  \citep{jullo07}. Unless otherwise noted, \lenstool\ models in this work assume the dPIE distribution, which has seven  parameters: centroid $x$, $y$; ellipticity $e$; position angle $\theta$; core radius $r_{\mathrm{core}}$; cut radius $r_{\mathrm{cut}}$; and normalization $\sigma_0$. Clusters are typically represented with one or more cluster-scale halos, supplemented with a number of galaxy-scale halos. Most of the parameters of cluster-scale halos are allowed to vary, with the exception of the cut radius, which for clusters is far outside the projected radius where it can be constrained by the lensing observables. If fixing the cut radius, each cluster-scale halo adds six free parameters to the lens model. To keep the number of overall optimized parameters manageable, it is common to assume that galaxy-scale halos of cluster member galaxies have similar locations, ellipticities, and position angles as the stellar mass distribution of the galaxies they represent, and that their mass and slope parameters can be linked to the luminosity of each galaxy through the parameterized \cite{faberjackson1976} scaling relations:

\begin{equation*}
\\
\begin{cases}
\referee{\sigma = \sigma^*\Big(\tfrac{L}{L^*}\Big)^{1/4}} \\
\referee{r_{\mathrm{core}} = r^*_{\mathrm{core}}\Big(\tfrac{L}{L^*}\Big)^{1/2}}\\
\referee{r_{\mathrm{cut}} =  r^*_{\mathrm{cut}}\Big(\tfrac{L}{L^*}\Big)^{1/2}}
\end{cases}  \\
\end{equation*}

\noindent This scaling adds two free parameters to the model: $r_{\mathrm{cut}}$ and $\sigma_0$. \referee{$r_\mathrm{core}$ has a minimal impact on mass models and is thus fixed to $0.15$ kpc \citep[][]{limousin2005, limousin2007, Eliasdottir2007}.}

Lens modeling is typically done iteratively, starting with a simple lens model and the most obvious sets of multiple images. Preliminary lens models can be used to assist in the identification and confirmation of further \referee{lensed sources}, which in turn are included as constraints in the next iteration. With more constraints, the complexity and flexibility of the model can be increased. The process continues until the model converges on a stable solution, where newly identified \referee{sources} are reliably predicted by the model and/or adding complexity does not significantly improve the goodness of fit or change the model output that is being sought (e.g., projected mass density, magnification, time delay, etc).
Modeling choices such as the number of halos are continuously evaluated, especially when new evidence is obtained, such as newly measured spectroscopic redshifts. 
\purple{Unless stated otherwise, all \lenstool\ models presented in this paper are optimized in the image plane and are run for at least 5000 steps past burn-in.}  All models are constructed using a standard fiducial positional uncertainty in the field between $0\farcs2$ and $0\farcs5$. \referee{The fiducial positional uncertainty is selected for each model based on the number of constraints in the model, where models with a high number of individual constraints but a low number of spectroscopic redshifts (i.e., models with many lensed source galaxies whose redshifts are optimized within the model) are given more tolerance in the optimization procedure. This uncertainty does not reflect the accuracy of the astrometry seen in the JWST imaging, which is incredibly small; it instead reflects the scatter between predicted and observed images typical of parametric lens models \citep{limousin2025}}.

\subsection{\wslap}\label{sec:wslap}
%----------------------------------------
This code was first described in \cite{Diego2005} and later expanded in \cite{Diego2007} to include weak lensing measurements as additional constraints.  A further development is presented in \cite{sendra2014}, where the code migrated from its native free-form nature to a hybrid type of modeling, where prominent cluster member galaxies are added into the lens model with a mass distribution that matches the observed light distribution.  

\wslap\,  places Gaussians in a predetermined grid of positions in the lens plane.  Each Gaussian contributes to the deflection field an amount that is proportional to its mass. This mass is optimized by the algorithm. A contribution from member galaxies is also precomputed by adopting a fiducial mass for them, that is later also optimized. A joint solution for the masses of the Gaussians, the renormalization of the mass of the member galaxies, and the unknown position of the lensed galaxies in the source plane is obtained by solving a system of linear equations. 
\begin{equation}
\Phi = \mathbf{\Gamma} X,
\label{Eq_SLWL_matrix_simple}
\end{equation}
where $\Phi$ is an array containing the observed positions of the arcs, $\Gamma$ is a known matrix, and $X$ is the vector with all the unknowns: masses in the Gaussian decomposition ($M$), multiplicative factors for the fiducial mass of the member galaxies ($C$), and source positions ($\beta_x$ and $\beta_y$). The solution is obtained using a quadratic programming optimization algorithm with the constraint $X>0$; we refer the reader to \cite{Diego2007} for full details on the optimization algorithm. \\

\subsection{Selection of Cluster Member Galaxies} \label{sec.clustermemberselection}
Cluster member galaxies were selected photometrically, using the red-sequence technique \cite{gladdersyee2000}. We generated a photometry catalogue with Source Extractor \citep{Bertin1996} in two filters that bracket the 4000\AA\ break feature in typical elliptical galaxies at the cluster redshift. Galaxies were identified as cluster members based on their color with respect to the red sequence in a color-magnitude diagram. In most clusters, we used archival \hst~data for both filters; the specific filter selection changes based on the availability of archival data and the cluster redshift (see \purple{\autoref{tab.redseq} for complete information on the red sequence construction}). The exception is SPT-CL\,J0516$-$5755, for which only one archival \hst~filter is available. For this cluster, we used the F150W2 as the red filter, as the 4000\AA\ break feature in this filter is comfortably red of F606W at the cluster redshift ($z=0.9656$). 

%%%%%%%%%%%%%%%%%%%%%%%%%%%%%%%%%%%%%%%%%%%
%CLUSTER LENSING NUMBER COUNTS
%%%%%%%%%%%%%%%%%%%%%%%%%%%%%%%%%%%%%%%%%%%

\begin{deluxetable}{lccc}[t!]
\tablecaption{\purple{Cluster Red Sequence Catalogue}\label{tab.redseq}}
\tablehead{
    \colhead{Cluster} & 
    \colhead{Bands} & 
    \colhead{Scatter [$\sigma$]} & 
}
\startdata
Abell\,68    & F110W/F814W    & 0.048 \\
RXC\,J0232.2$-$4420    & F105W/F814W    & 0.031  \\
SMACS\,J2031.8$-$4036    &F814W/F606W    & 0.034  \\
MACS\,J0027.8+2616    & F775W/F606W   & 0.120 \\
RXC\,J2211.7$-$0349    & F814W/F606W    & 0.029 \\
MACS\,J0553.4$-$3342    & F814W/F606W    & 0.035 \\
MACS\,J0358.8$-$2955    & F814W/F606W   & 0.043  \\
MACS\,J1621.4+3810    & F814W/F606W   & 0.027  \\
RCS2\,032727$-$132623    & F814W/F606W    & 0.061  \\
MACS\,J2129.4$-$0741    & F814W/F606W    & 0.055 \\
PSZ1\,G091.83+26.11    & F110W/F814W    & 0.044   \\
RX\,J0152.7$-$1357    & F160W/F625W   & 0.104   \\
SPT$-$CL\,J0516$-$5755    & F150W2/F606W    & 0.176 \\
SPT$-$CL\,J2011$-$5228    & F125W/F606W    & 0.121 \\
\enddata
\tablecomments{ Column (1): the cluster. Column (2): the bands used to create the cluster member catalog for each cluster are listed as pairs. Column (3): the scatter around the red sequence.}

\end{deluxetable}

%%%%%%%%%%%%%%%%%%%%%%%%%%%%%%%%%%%%%%%%%%%
%%%%%%%%%%%%%%%%%%%%%%%%%%%%%%%%%%%%%%%%%%%
\subsection{Cluster Models}\label{sec.clustermodels}
\begin{figure*}[htbp!] 
    \begin{minipage}{1\textwidth}
   \centering 
    \includegraphics[width=0.76\linewidth]{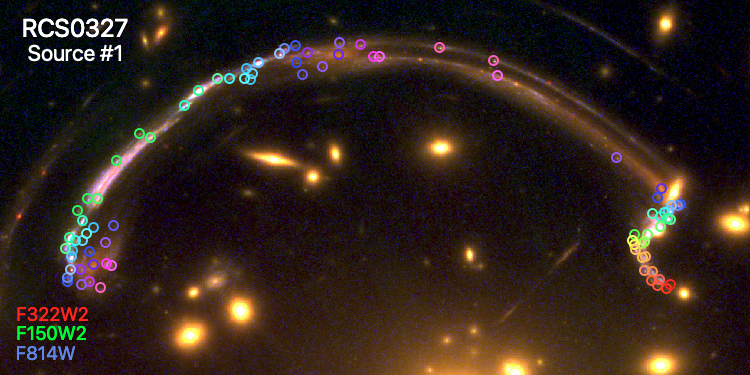}
        \includegraphics[width=0.38\linewidth]{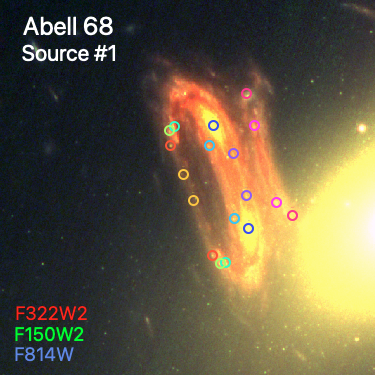}
        \includegraphics[width=0.38\linewidth]{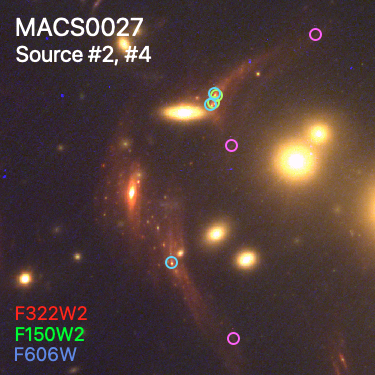}        
    \includegraphics[width=0.38\linewidth]{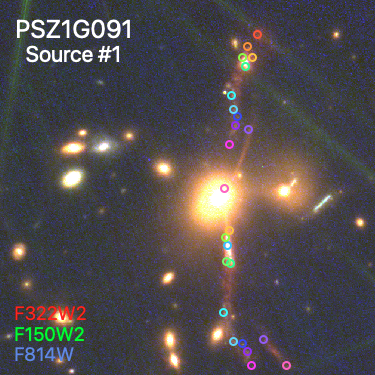}
    \includegraphics[width=0.38\linewidth]{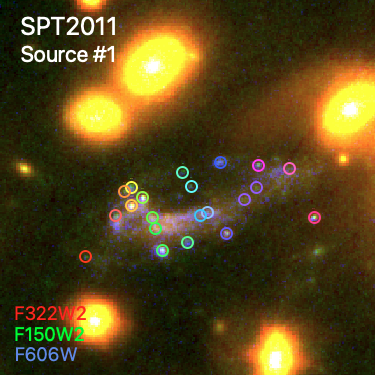}
        \caption{Snapshots of five lensed images from the SLICE program, including the primary tangential arc from Source  \#1 in RCS 0327 (top), \referee{Source} \#1 in Abell\,68 (middle left), \referee{Sources} \#2 and \#4 from MACS0027 (middle right), \referee{Source} \#1 in PSZ1G091 (bottom left), and the radial arc from \referee{Source} \#1 in SPT2011 (bottom right). The constraints identified in the process of constructing the lens models are color-coded but are not labeled for visual clarity. Some constraints shown in this figure include candidate clumps that were not used in the construction of the final lens model presented in this paper. The complete constraint information is provided in \autoref{tab.MultipleImages}. The large number of identified emission knots \referee{(clumps)} in these lensed galaxies is possible due to the high-resolution infrared imaging capabilities of \jwst. Each additional constraint imposes more precise limits on the modeled mass distribution of these clusters, increasing the potential of lens modeling to explore the fundamental properties of clusters.  
        }\label{fig.arcsnapshots}
    \end{minipage}
\end{figure*}

In the following sections, we present details for each of the 14 clusters modeled in this paper. Images of the clusters and the strong lensing constraints are shown in~\autoref{fig.allmodels}.  In the description for \purple{each model}, \referee{we specify the number of unique lensed source galaxies}. The \referee{``}number of images\referee{"} is the total number of multiple images \referee{of all lensed galaxies (not including clumps within the galaxies)}. Where clumps within a lensed \referee{galaxy} (e.g.~\autoref{fig.arcsnapshots}) are used to constrain the model, the \referee{``}number of \referee{positional} constraints\referee{" accounts for} multiple images of the individual clumps. Otherwise, the number of \referee{positional} constraints equals the number of images \referee{of lensed galaxies} used in the model. \purple{As an example, suppose \referee{we identify three lensed source galaxies in a cluster field}. \referee{Source} 1 has three multiple images and does not use any clumps as constraints. \referee{Source} 2 has five multiple images and uses four clumps per image as constraints. \referee{Source} 3 has three multiple images and uses two clumps per image as constraints. In this hypothetical cluster, the total number of \textit{\referee{lensed source galaxies}} is 3, the total number of \textit{images} is 11 \referee{($3+5+3$)}, and the total number of \textit{\referee{positional} constraints}\footnote{We note that each positional constraint is a 2D vector, but we count each image of a lensed source (galaxy or clump) as a single position.} is 29 \referee{($3 + 5\times4 + 3\times2$)}. If no clumps were used as constraints in \referee{Source} 2 or \referee{Source} 3, then the total number of \textit{images} would be 11 \referee{(3+5+3)}, and the total number of \textit{\referee{positional} constraints} would also be 11 \referee{(3+5+3)}. \autoref{tab.clusterstats} contains a summary of all the \referee{lensed source galaxies} used as constraints in the models.}~

\autoref{tab.clustermass} provides mass estimates and Einstein radius measurements from the \lenstool~models. The cluster best-fit \purple{and median} parameters for each of the potentials and parameters in the \lenstool~models are given in~\autoref{tab.bestfitparams}. \purple{The reported errors are calculated from the $68\%$ confidence interval for each parameter, which are optimized using a uniform prior. As a result, some error distributions are asymmetric.} A complete list of the arcs and model redshifts for each cluster is given in \autoref{tab.MultipleImages}.~ Candidate \referee{sources} and arcs are not included in the count if they were not used as lensing constraints, but are shown in the figures and \purple{\autoref{tab.MultipleImages}} for reference.
All the lens models presented here are publicly available at the Strong Lensing Cluster Atlas Data Base\footnote{the
\href{https://data.lam.fr/sl-cluster-atlas/}{Strong Lensing Cluster Atlas Data Base} is hosted at Laboratoire d’Astrophysique de Marseille.
}, which contains mass maps (e.g. shear, convergence, amplification), as well as parameter files used for the lens modeling (e.g. lists of constraints,  cluster member catalogues).

%%%%%%%%%%%%%%%%%%%%%%%%%%%%%%%%%%%%%%
%CLUSTER LENS MODEL FIGURE
%%%%%%%%%%%%%%%%%%%%%%%%%%%%%%%%%%%%%%

\begin{figure*}
 \centering
 \begin{minipage}{0.9\textwidth}  
        \gridline{
            \fig{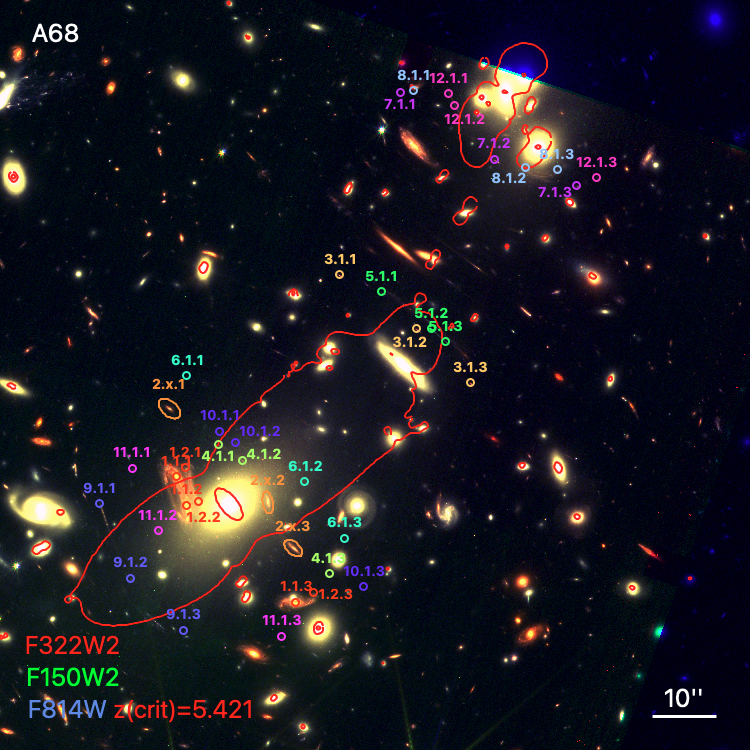}{0.5\textwidth}{}
            \fig{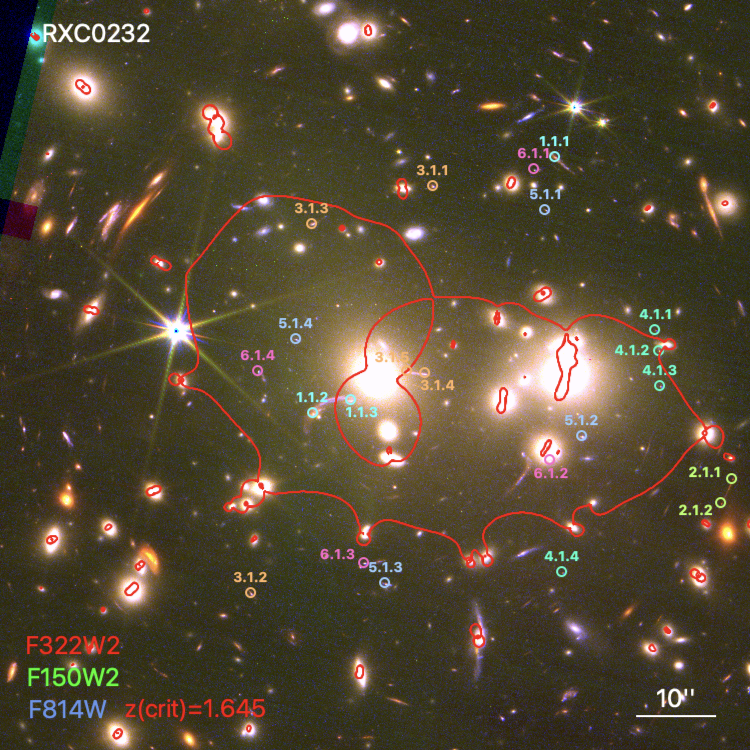}{0.5\textwidth}{}}
        \vspace*{-9mm}
        \gridline{
            \fig{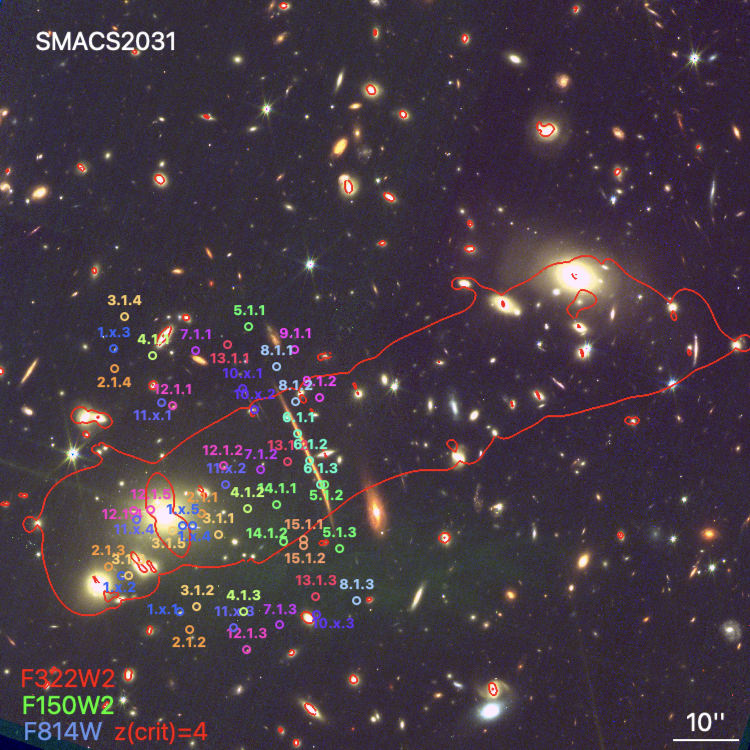}{0.5\textwidth}{}
            \fig{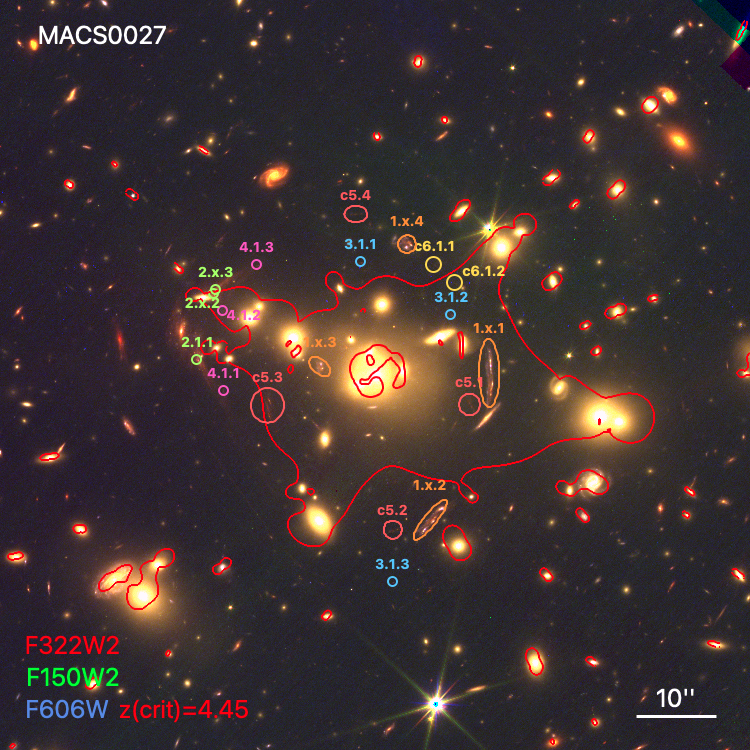}{0.5\textwidth}{}}
        \vspace*{-9mm}
        \gridline{
            \fig{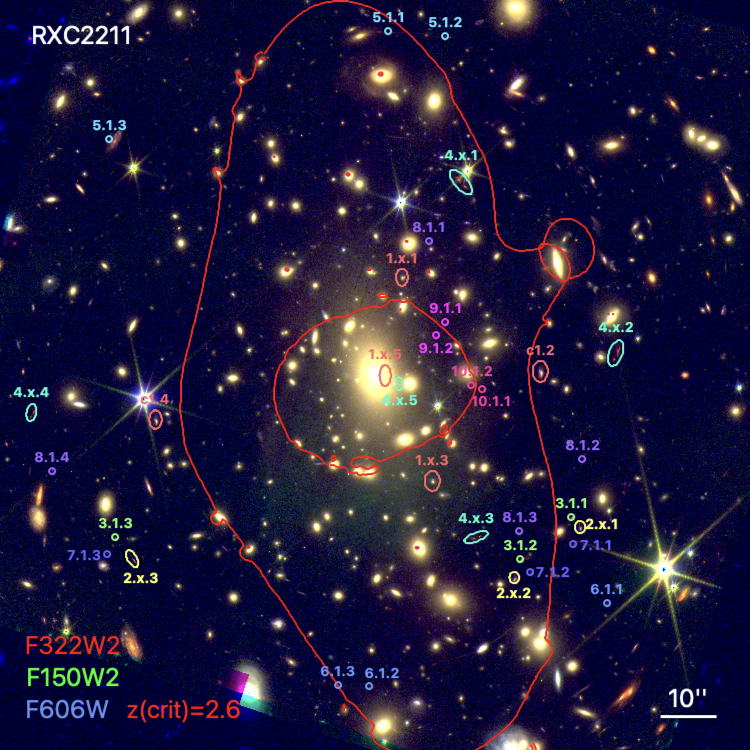}{0.5\textwidth}{}
            \fig{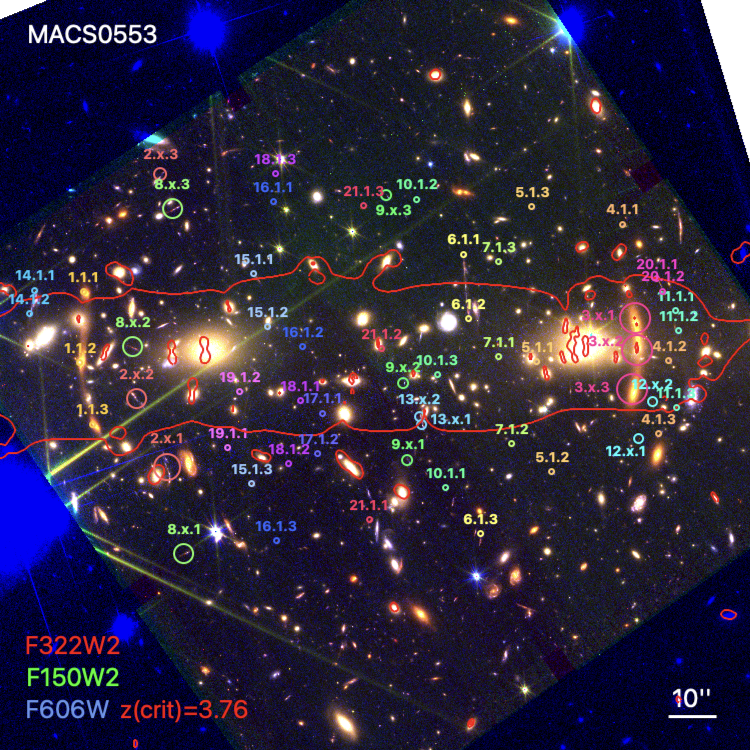}{0.5\textwidth}{}}
            \vspace*{-9mm}
    \caption{\textit{continued}}
    \end{minipage}
\end{figure*}

\begin{figure*}
\centering
\addtocounter{figure}{-1}
 \begin{minipage}{0.9\textwidth}  
        \gridline{
            \fig{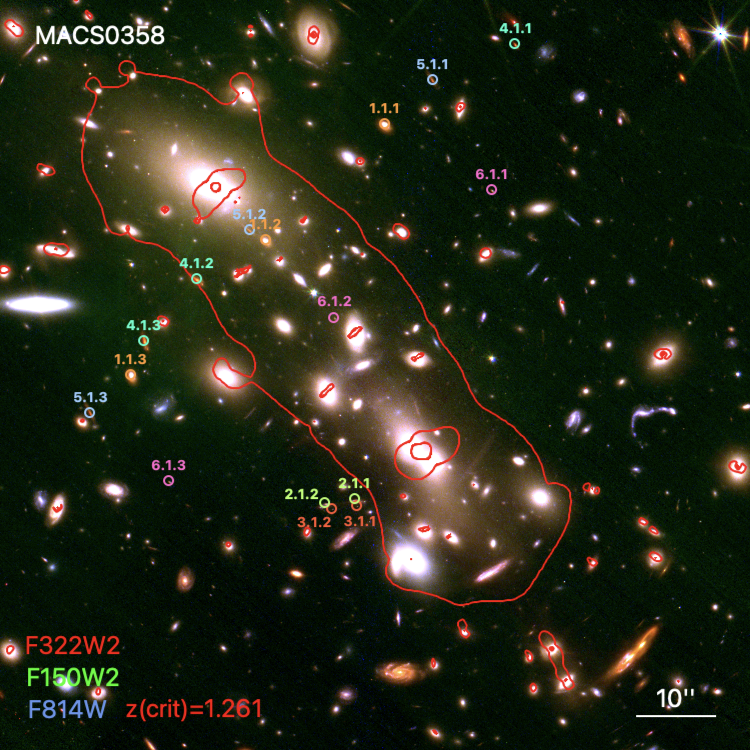}{0.5\textwidth}{}
            \fig{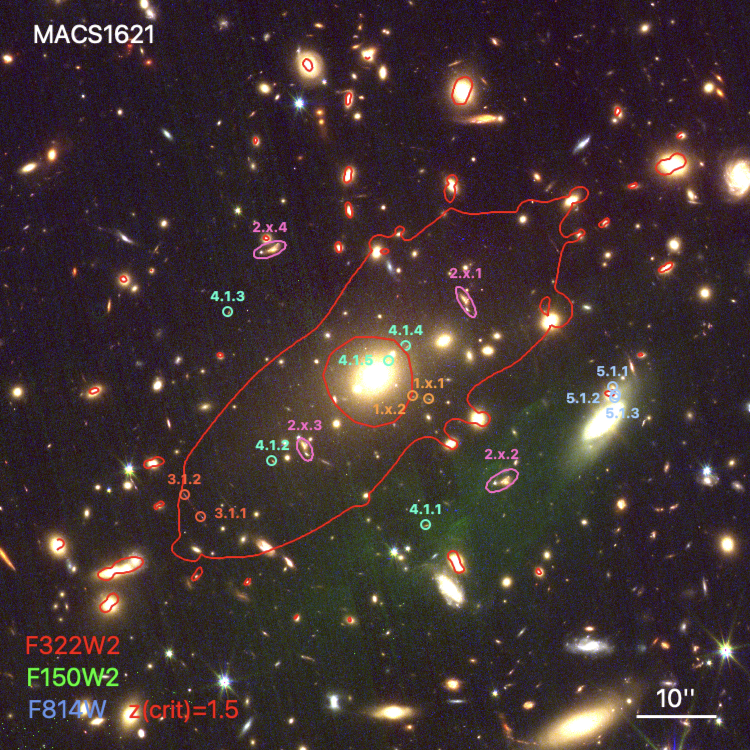}{0.5\textwidth}{}}
        \vspace*{-9mm}
        \gridline{
            \fig{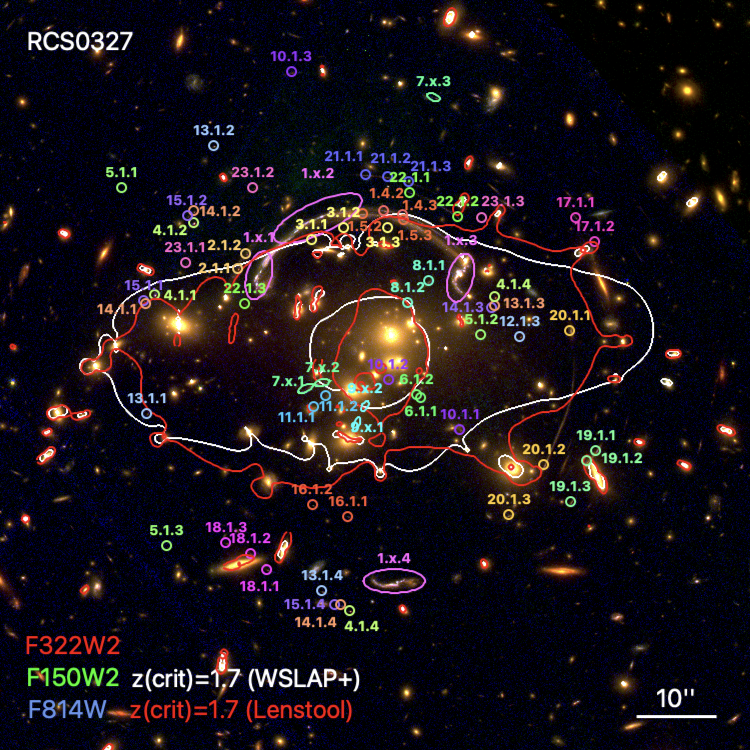}{0.5\textwidth}{}
            \fig{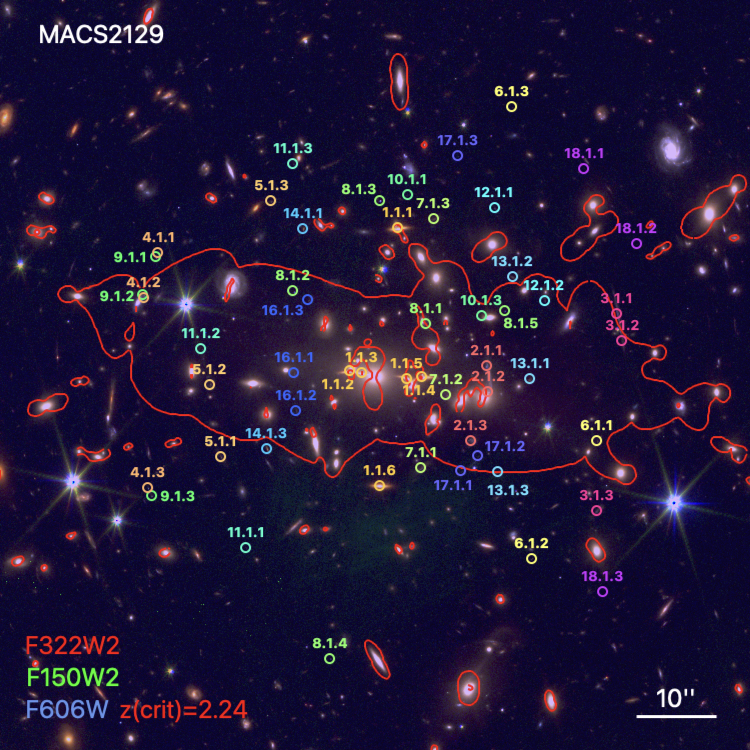}{0.5\textwidth}{}}
        \vspace*{-9mm}
        \gridline{
            \fig{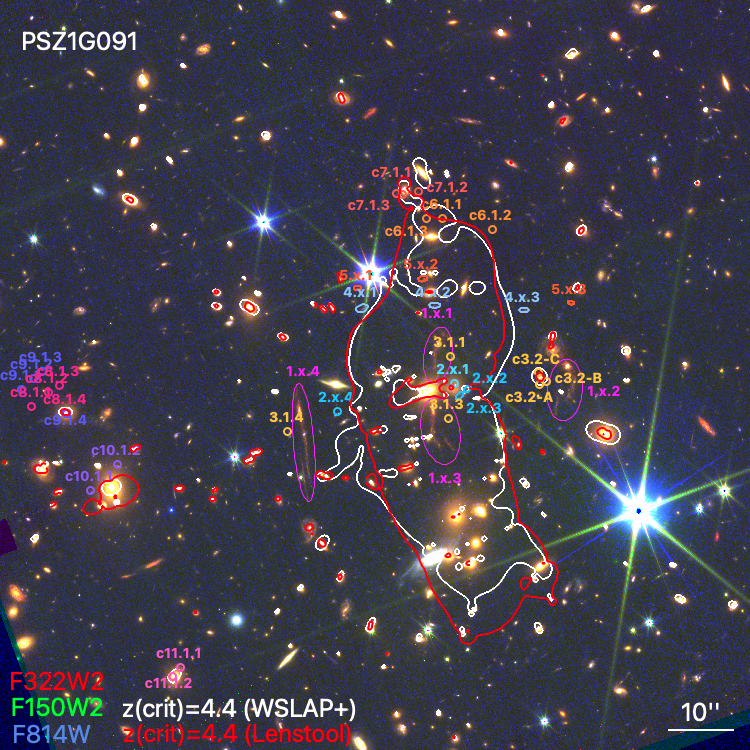}{0.5\textwidth}{}
            \fig{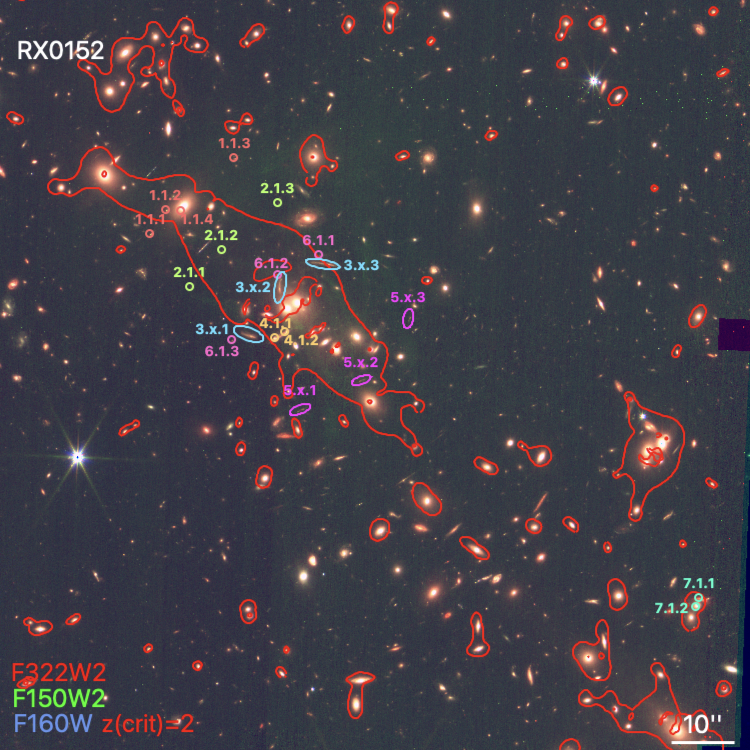}{0.5\textwidth}{}}
        \vspace*{-9mm}
    \caption{\textit{continued}}
    \end{minipage}
\end{figure*}

\begin{figure*}[htbp]
\centering
\addtocounter{figure}{-1}
\begin{minipage}{0.9\textwidth}
    \gridline{
        \fig{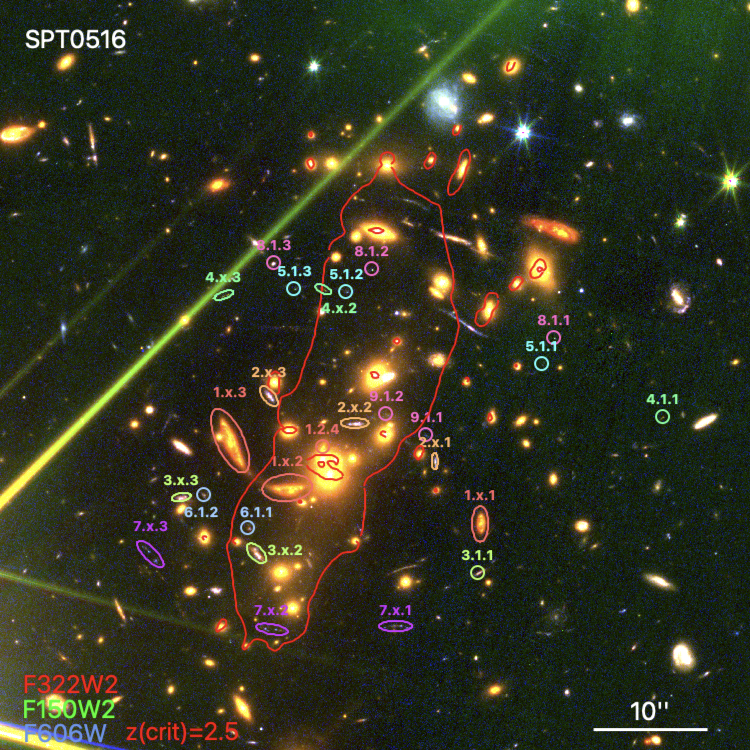}{0.5\textwidth}{}
        \fig{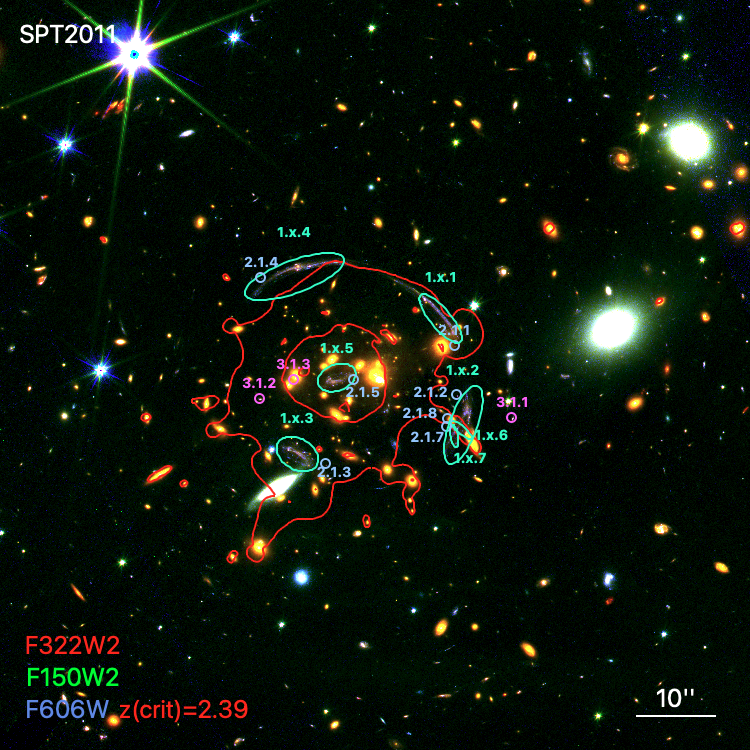}{0.5\textwidth}{}}
        \vspace*{-7mm}
        \caption{False color images of all clusters modeled in this paper. All images are oriented North up and East to the left. The color composites are rendered from the two SLICE \jwst~NIRCam bands and one of the available archival \hst~bands in the blue channel, as listed in the bottom left of each panel. We overplot the best-fit critical curves for a source at the redshift listed in each panel. Lens models constructed with \lenstool~have their critical curves plotted in red, while lens models constructed with \wslap~have their critical curves plotted in white.  Sets of multiple images of lensed sources are color-coded and labeled. To reduce clutter, we label crowded individual clumps as `[\referee{Source}\#].x.[Image\#]', where x refers to the clumps. Where applicable, candidate \referee{lensed sources} are labeled using the letter ``c'', e.g. ``c1.1.1'' indicates a candidate \referee{source} that is not included in the optimization of the lens model. Two \referee{lensed source galaxies} in PSZ1\,G091.83+26.11 are outside the field of view, but are included in the model. See \autoref{tab.MultipleImages} for a complete list of all the images and clumps for the \referee{lensed source galaxies} in each cluster.
        }\label{fig.allmodels}
\end{minipage}
\end{figure*}

%%%%%%%%%%%%%%%%%%%%%%%%%%%%%%%%%%%%%%
%%%%%%%%%%%%%%%%%%%%%%%%%%%%%%%%%%%%%%

%%%%%%%%%%%%%%%%%%%%%%%%%%%%%%%%%%%%%%%%%%%
%CLUSTER LENSING NUMBER COUNTS
%%%%%%%%%%%%%%%%%%%%%%%%%%%%%%%%%%%%%%%%%%%

\begin{deluxetable*}{lccccccc}[t!]
\tablecaption{Cluster Statistics Table\label{tab.clusterstats}}
\tablehead{
    \colhead{Cluster} & 
    \colhead{\# \referee{Lensed}} & 
    \colhead{\# Images$^a$} & 
    \colhead{\# \referee{Positional}} &
    \colhead{\# Tangential Arcs} & 
    \colhead{\# Radial Arcs} & 
    \colhead{\# $z_{spec}$} & 
    \colhead{Model RMS ($''$)} \\
    \colhead{} & 
    \colhead{ }{\referee{Galaxies}} & 
    \colhead{} & 
    \colhead{ }{\referee{Constraints$^b$}} &
    \colhead{} & 
    \colhead{} & 
    \colhead{} & 
    \colhead{} 
}
\startdata
Abell\,68    &12    &36    &42    &36    &0    &5    & \purple{$0\farcs44$} \\
RXC\,J0232.2$-$4420    &7    &26    &26    &18    &4    &2    & \purple{$0\farcs43$} \\
SMACS\,J2031.8$-$4036    &15    &53    &67    &49    &4    &13    &$0\farcs40$\\
MACS\,J0027.8+2616    &4    &19    &31    &19    &0    &2    &\purple{$0\farcs48$}\\
RXC\,J2211.7$-$0349    &10    &32    &45    &28    &4    &1    & \purple{$0\farcs57$}\\
MACS\,J0553.4$-$3342    &21    &57    &102    &57    &0    &4    & \purple{$0\farcs66$}\\
MACS\,J0358.8$-$2955    &6    &16    &16    &16    &0    &2    & \purple{$0\farcs21$} \\
MACS\,J1621.4+3810    &5    &16    &22    &12    &4    &0    & \purple{$0\farcs26$} \\
RCS2\,032727$-$132623    &23    &61    &184    &47    &14    &5    & \purple{$0\farcs77$} \\
MACS\,J2129.4$-$0741    &17    &53    &53    &49    &4    &13    &$0\farcs79$\\
PSZ1\,G091.83+26.11    &5    &\purple{19}    &89    &\purple{16}    &\purple{3}    &0    & $0\farcs47$ \\
RX\,J0152.7$-$1357    & 7    & 20    &26    &20    &0    &1    &$0\farcs45$\\
SPT$-$CL\,J0516$-$5755    &10    &25    &53    &25    &0    &0    & \purple{$0\farcs21$}\\
SPT$-$CL\,J2011$-$5228    &3    &14    & 96    &12    &2    &1    & \purple{$0\farcs33$} \\
\enddata
\tablecomments{Number counts for various properties of the lens model presented in this paper. 
\referee{$^a$The \# Images column sums the multiple images of lensed galaxies (not including clumps). 
$^b$The \# Positional Constraints sums the multiple images of all lensed features (including clumps) that were used to constrain the strong lensing model.} }

\end{deluxetable*}

%%%%%%%%%%%%%%%%%%%%%%%%%%%%%%%%%%%%%%%%%%%
%MASS MODELING MEASUREMENTS 
%%%%%%%%%%%%%%%%%%%%%%%%%%%%%%%%%%%%%%%%%%%

\begin{deluxetable*}{lcccc}[htbp!]
\tablecaption{Cluster Mass Table ($10^{14}\ {\rm M}_{\odot}$ with \purple{$68\%$ confidence interval} uncertainty)\label{tab.clustermass}}
\tablecolumns{5}
\tablehead{
    \colhead{Cluster} & 
    \colhead{ M($\textless200 $kpc) } & 
    \colhead{ M($\textless500$ kpc) } & 
    \colhead{$R_E$ ($''$)} & 
    \colhead{$R_E$ ($''$)}  \\[-8pt]
    \colhead{} &
    \colhead{}     & 
    \colhead{}    & 
    \colhead{$z=2$}       & 
    \colhead{$z=9$}       
}
\startdata
Abell\,68               & \purple{ $1.60^{+0.07}_{-0.26}$ }  & \purple{ $4.90^{+0.37}_{-0.81}$ }   & \purple{ $16.38_{-0.19}^{+0.38}$  } & \purple{ $19.72_{-0.48}^{+0.43}$                 } \\ 
RXC\,J0232.2$-$4420     & \purple{ $1.85^{+0.05}_{-0.04}$ }  & \purple{ $4.51^{+0.21}_{-0.21}$ }   & \purple{ $25.23_{-0.75}^{+0.38}$  } & \purple{ $29.53_{-0.99}^{+0.41}$  } \\
SMACS\,J2031.8$-$4036   & \purple{ $1.38^{+0.02}_{-0.01}$ }  & \purple{ $4.47^{+0.21}_{-0.10}$ }   & \purple{ $21.70_{-0.86}^{+0.23} $ } & \purple{ $28.12_{-0.48}^{+0.10}$  } \\ 
MACS\,J0027.8+2616      & \purple{ $1.73^{+0.07}_{-0.18}$ }  & \purple{ $5.09^{+0.37}_{-0.84}$ }   & \purple{ $10.90_{-1.08}^{+1.21}$  } & \purple{ $18.65_{-2.44}^{+0.54}$  } \\ 
RXC\,J2211.7$-$0349     & \purple{ $2.81^{+0.02}_{-0.01}$ }  & \purple{ $7.04^{+0.11}_{-0.08}$ }   & \purple{ $43.98^{+0.61}_{-0.55}$  } & \purple{ $53.35^{+0.98}_{-0.91}$  } \\ 
MACS\,J0553.4$-$3342    & \purple{ $1.84^{+0.00}_{-0.01}$ }  & \purple{ $6.15^{+0.03}_{-0.04}$ }   & \purple{ $32.33_{-0.10}^{+0.09}$  } & \purple{ $41.98_{-0.06}^{+0.11}$  } \\ 
MACS\,J0358.8$-$2955    & \purple{ $2.15^{+0.14}_{-0.03}$ }  & \purple{ $5.99^{+1.38}_{-0.11}$ }   & \purple{ $16.14_{-2.21}^{+6.96}$} & \purple{ $32.78_{-0.86}^{+1.62}$ } \\ 
MACS\,J1621.4+3810      & \purple{ $1.62^{+0.07}_{-0.06}$ }  & \purple{ $3.91^{+0.26}_{-0.29}$ }   & \purple{ $21.59_{-0.69}^{+0.44}$  } & \purple{ $27.32_{-1.40}^{+0.01}$  } \\ 
RCS2\,032727$-$132623   & \purple{ $1.91^{+0.28}_{-0.29}$ }  & \purple{ $5.14^{+0.68}_{-0.82}$ }   & \purple{ $23.60_{-0.25}^{+0.05}$  } & \purple{ $31.86_{-0.31}^{+0.20}$  } \\ 
MACS\,J2129.4$-$0741    & \purple{ $1.84^{+0.01}_{-0.03}$ }  & \purple{ $5.32^{+0.10}_{-0.19}$ }   & \purple{ $20.80_{-0.00}^{+0.41}$   } & \purple{ $29.77_{-0.00}^{+0.71}$  } \\ 
PSZ1\,G091.83+26.11     & \purple{ $1.89^{+0.02}_{-0.00}$ }  & \purple{ $7.61^{+0.21}_{-0.12}$ }   & \purple{ $4.64_{-0.37}^{+0.00}$   } & \purple{ $24.82_{-0.31}^{+0.19}$  } \\ 
RX\,J0152.7$-$1357      & \purple{ $1.14^{+0.13}_{-0.10}$ }  & \purple{ $3.44^{+0.40}_{-0.42}$ }   & \purple{ $11.15_{-0.90}^{+1.75}$  } & \purple{ $17.16_{-1.44}^{+5.35}$  } \\ 
SPT-CL\,J0516$-$5755    & \purple{ $1.74^{+0.08}_{-0.04}$ }  & \purple{ $5.10^{+0.42}_{-0.18}$ }   & \purple{ $7.23_{-2.05}^{+0.35}$   } & \purple{ $21.51_{-2.57}^{+0.77}$  } \\
SPT-CL\,J2011$-$5228    & \purple{ $2.51^{+0.22}_{-0.00}$ }  & \purple{ $7.73^{+1.75}_{-0.00}$ }   & \purple{ $8.94_{-1.06}^{+0.04}$   } & \purple{ $33.48_{-3.08}^{+0.47}$  } \\  
\enddata
\tablecomments{Summary of mass estimates from all SL models. Columns (2) and (3): the masses enclosed within 200 kpc and within 500 kpc, respectively. Columns (3) and(4): measurements of the cluster’s Einstein radius $R_E$ at two different redshifts. For A68, the measurement of $R_E$ is limited to the major cluster component, to ensure consistency between the two redshift measurements. The mass estimates are calculated starting from the reference coordinates listed in Column (2) of~\autoref{tab.bestfitparams}. All uncertainties are calculated relative to the best-fit model with an asymmetric 68\% confidence level.}

\end{deluxetable*}
%%%%%%%%%%%%%%%%%%%%%%%%%%%%%%%%%%%%%%%%%%%
%%%%%%%%%%%%%%%%%%%%%%%%%%%%%%%%%%%%%%%%%%%

%%%%%%%%%%%%%%%%%%%%%%%%%%%%%%%%%%%%
%%%%%%%%%%%%%%%%%%%%%%%%%%%%%%%%%%%%

%%%%%%%%%%%%%%%%%%%%%%%%%%%%%%%%%%%%
%%%% CLUSTER MASS FIGURE
\begin{figure*}[h!] 
   \centering 
        \includegraphics[width=1\linewidth,height=1\linewidth]{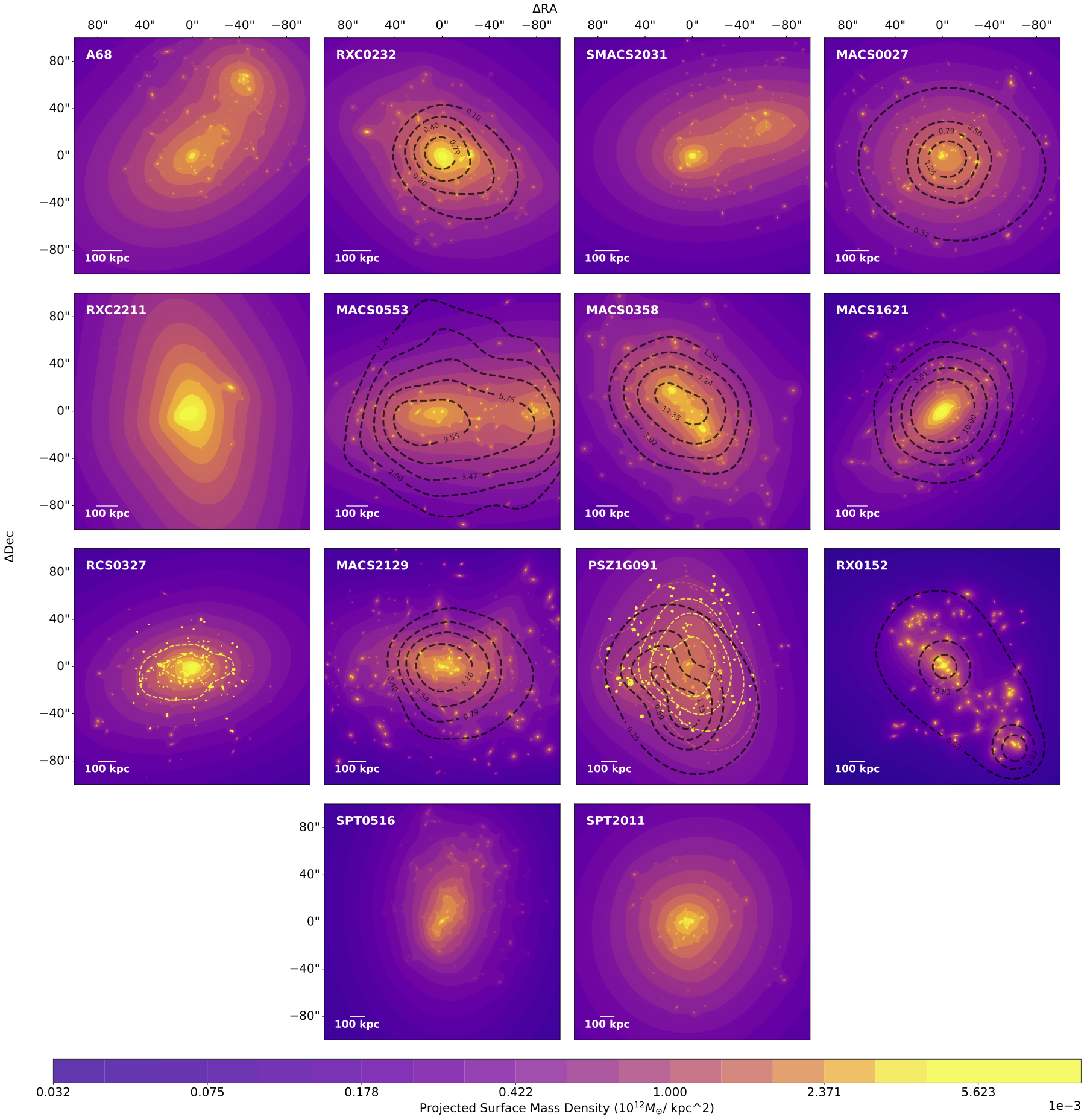}
        \caption{\label{fig.projmass} The total projected surface mass density of each cluster is measured within a 200\arcsec$\times$200\arcsec\ box centered on the reference coordinates for each lens model made with \lenstool, which are listed in the second column of \autoref{tab.bestfitparams}. \purple{The \lenstool\ mass density map is shown as logarithmically-spaced contours between $10^8$ $M_\odot$ and $10^{11}$ $M_\odot$, which are spaced according to the scale given in the colorbar. Mass contours across the grids used in the two \wslap\ models are shown as the dashed lines and are colored according to the same scale as the \lenstool\ maps.} Chandra X-ray observations are available for eight out of the fourteen clusters. For the purposes of presentation, we smooth the X-ray map with the \texttt{csmooth} routine using adaptive smoothing to $2.5\sigma$ significance. We convolve the smoothed X-ray map with a Gaussian of standard deviation 5\,pixels and overplot logarithmically-spaced contours in black dashed lines over the projected mass map\purple{, with contour levels given explicitly in the plot}. A uniform scale length of 100\,kpc is shown in the bottom left corner of each plot.  A high-quality version of this figure is available in the published journal version of this manuscript. }
\end{figure*} 
%%%%%%%%%%%%%%%%%%%%%%%%%%%%%%%%%%%%

\subsubsection{Abell\,68} 

Abell\,68 (A68), located at $z=0.2546$, is optimized with a total of 12 lensed source galaxies (36 images and 42 lensing constraints), where 5 \referee{source galaxies} have spectroscopic redshifts and 7 unknown redshifts are optimized within the model. A total of 206 cluster member galaxies are fitted with a scaling relation \purple{(as described in ~\autoref{sec.clustermemberselection})} optimizing 2 parameters. \purple{The reference luminosity $L^*$ used in this relation is from the BCG of the cluster (F814W = 17.94).} The model uses four potentials to optimize the mass of the cluster. The final model has an rms of \purple{ $0\farcs44$} , and the mass enclosed within 200\,kpc and 500 kpc is \purple{ $\sim1.60\times10^{14}M_{\odot}$ and $\sim4.90\times10^{14}M_{\odot}$}, respectively. We provide a detailed description of the cluster modeling in the following paragraph. 

A68 contains a wealth of lensed galaxies that are scattered between the two ends of its bimodal light distribution. We update the model presented in \cite{richard2007} by doubling the number of constraints present in the model (from 21 to 42), which we accomplish by confirming candidates from \cite{richard2007}, such as the candidate \referee{Source} C10 and the candidate lensing images C2c and C23c, which are designated in this paper as \referee{Sources} \#6 and \#4, respectively. We add a total of nine new \referee{lensed source galaxies} into the model. The resolution of the \jwst~imaging also reveals a great deal of substructure in the main lensed dusty galaxy in this cluster, which is here designated as \referee{Source}\#1. We are able to identify at least two different multiply-imaged clumps within both the dual radial images and the third counter-image of this galaxy, which we incorporate into the model as constraints. This increases the total number of lensing constraints present within a single lensed image. We utilize the spectroscopic redshift measurements presented in \cite{richard2007} for the five previously-identified \referee{sources} in this model. Reproducing all of the lensing constraints across the breadth of the cluster requires the inclusion of three cluster-scale DM halos into the model: one located at the BCG, one located in the northern portion of the cluster, and one located near the second-brightest galaxy in the cluster. We also separately optimize the parameters of the galaxy located next to the first image of \referee{Source} \#1, as well as the BCG. The model redshift predictions for the sources used as constraints in the model are under $z\sim4$ in all cases.

%%%%%%%%%%%%%%%%%%%%%%%%%%%%%%%%%%%%%%%%%%%
%%%%%%%%%%%%%%%%%%%%%%%%%%%%%%%%%%%%%%%%%%%

\subsubsection{RXC\,J0232.2$-$4420} 
RXC\,J0232.2$-$4420 (RXC0232), located at $z=0.2836$, is optimized with a total of 7 \referee{lensed source galaxies} (26 multiple images and 26 lensing constraints), where 2 \referee{source galaxies} have spectroscopic redshifts and 5 redshifts are optimized within the model. A total of 115 cluster member galaxies \purple{ are fitted with a scaling relation based on a reference luminosity $L^*$ for the BCG of the cluster (F814W = 18.23).} The model uses 5 potentials to describe the mass of the cluster. The final model has an rms of \purple{$0\farcs43$}, and the mass enclosed within 200\,kpc and 500\,kpc is \purple{$\sim1.85\times10^{14}M_{\odot}$ and $\sim4.51\times10^{14}M_{\odot}$}, respectively. We provide a detailed description of the cluster modeling in the following paragraph. 

RXC0232 \purple{was observed as part} of the RELICS cluster program \citep{coe2019} and displays a bimodal light distribution with two BCGs ($\mathrm{BCG_A}$ and $\mathrm{BCG_B}$) separated by 24\arcsec\ along the East-West direction. Both the X-ray and radio contours are centered on $\mathrm{BCG_A}$ in the East \citep{kale2019}. We report six \referee{lensed source galaxies} around these two BCGs (two radial \referee{multiple images}), one of them being at a spectroscopic redshift $z_{spec}=1.645$ \citep{ramesh2019}. We report one \referee{lensed source galaxy} at a spectroscopic redshift $z_{spec}=1.467$ \citep{ramesh2019} located further East, just outside the \jwst~field of view but observed thanks to the \hst~imaging. It is located around a group of three galaxies. The mass model includes two large-scale DM halos coincident with each BCG. The two BCGs are also optimized individually. A group-scale DM halo is included as well to account for the Eastern galaxy group. The central peak of the X-ray imaging is aligned with the center of the mass peak of the cluster (see \autoref{tab.xray} for the X-ray observation details and \autoref{fig.projmass} for the X-ray contours in cyan and projected mass contours in black). The model optimizes the fits for the redshifts of five \referee{lensed source galaxies}. The best-fit optimization of the redshift for \referee{Source} \#2 is \purple{$5.76^{+0.15}_{-0.86}$}, while the other four \referee{sources} have optimized redshifts under $z\sim3$. 

%%%%%%%%%%%%%%%%%%%%%%%%%%%%%%%%%%%%%%%%%%%
%%%%%%%%%%%%%%%%%%%%%%%%%%%%%%%%%%%%%%%%%%%

\subsubsection{SMACS\,J2031.8$-$4036} 
SMACS\,J2031.8$-$4036 (SMACS2031), located at $z=0.3416$, is optimized with a total of 15 \referee{lensed source galaxies} (53 multiple images and 67 lensing constraints), where 13 \referee{source galaxies} have spectroscopic redshifts and 2 redshifts are optimized within the model. A total of 165 cluster member galaxies \purple{ are fitted with a scaling relation based on a reference luminosity $L^*$ for the BCG of the cluster (F814W = 18.74).} The model uses three potentials to describe the mass of the cluster. The final model has an rms of $0\farcs40$, and the mass enclosed within 200\,kpc and 500\,kpc is $\sim1.38\times10^{14}M_{\odot}$ and $\sim4.47\times10^{14}M_{\odot}$, respectively. We provide a detailed description of the cluster modeling in the following paragraphs. 

The SLICE model of SMACS2031 is based on (and updates) the model presented in \citet{richard21} as part of the MUSE Cluster Atlas- a project targeting massive galaxy clusters with a combination of MUSE spectroscopy and \hst\ imaging. Based on the \hst\ data, SMACS2031 appears to be undergoing a major merger, with two distinct BCGs separated by 72\arcsec. The apparent merger axis is oriented at a position angle of 60\,degrees West of North. As a Cluster Atlas member, SMACS2031 has significant spectroscopic coverage, though only over its Eastern half, as the Western portion was not targeted by the survey. The \citet{richard21} model identified 13 multiply-imaged sources (consisting of 46 multiple images); however, due to the relatively shallow \hst\ data, several of these \referee{lensed source galaxies} are not detected in imaging, but only as emission line features in MUSE. In contrast, SLICE \jwst~imaging reveals clear continuum emission (in both F150W2 and F332W2) for all 13 \referee{sources}, improving the centroiding precision of each image by up to $\sim 0.5$\arcsec. The improved resolution of \jwst~also reveals complex morphologies in many of the \referee{lensed source galaxies}, with \referee{Sources} \#1, \#10, and \#11 each having identifiable clumps in their continuum appearances, allowing us to use each of these clumps as a distinct model constraint.  

In addition to confirming and updating the positions of existing \referee{lensed source galaxies and their associated clumps}, we also identify two new, faint, arc-like objects (which we label as \referee{Sources} \#14 and \#15) in the Eastern half of the cluster. Although we also detect several long, extended arc structures around the Western half of SMACS2031, these arcs have a distinctly smooth appearance, with no obvious counter-images, making it difficult to know if they are merging pairs or simply distorted single-image galaxies. Therefore, at present, we do not include any of the Western arcs as constraints, but this can be revisited with deeper imaging and/or spectroscopic follow-up.

Using these constraints, we construct a new lens model that includes three cluster- and galaxy-scale mass halos in the model (see \autoref{tab.bestfitparams}) -- two to represent the cluster-scale halos of the Eastern and Western components of SMACS2031, and a third galaxy-scale halo to account for the Eastern BCG. Our model is then complemented with 165 smaller-scale halos to account for cluster member galaxies. We specifically model the Eastern BCG separately from other galaxies (see~\autoref{sec.lenstool}) because it lies close to individual lensed galaxies and could bias the fit.

After optimizing the model, we find a best-fit with a rms of $0\farcs40$. The best-fit position of the Eastern large-scale cluster halo is nearly colinear with the BCG, while the Western halo is offset by $\sim$12\arcsec\ to the South (similar to the results of the \citealt{richard21} model). However, the \jwst~imaging of the Eastern BCG reveals that the galaxy has a significant diffuse/shell-like appearance, suggesting that it has recently been disrupted, and possibly offering clues as to the nature of the cluster halo offset. The lack of constraints around the Western BCG make it difficult to obtain a proper centroid, which may also account for this offset.
The model optimizes the redshifts of 2 \referee{lensed source galaxies}, \#14 and \#15. The best-fit optimization of \referee{Source} \#15 is $6.89^{+0.08}_{-0.20}$, while \referee{Source} \#14 is under $z\sim2$.  

%%%%%%%%%%%%%%%%%%%%%%%%%%%%%%%%%%%%%%%%%%%
%%%%%%%%%%%%%%%%%%%%%%%%%%%%%%%%%%%%%%%%%%%

\subsubsection{MACS\,J0027.8$+$2616} 
MACS\,J0027.8$+$2616 (MACS0027), located at $z=0.365$, is optimized with a total of 4 \referee{lensed source galaxies} (19 multiple images and 31 lensing constraints), where 2 \referee{source galaxies} have spectroscopic redshifts and 2 redshifts are optimized within the model. A total of 170 cluster member galaxies \purple{ are fitted with a scaling relation based on a reference luminosity $L^*$ for the BCG of the cluster (F606W = 19.39).} The model uses one potential to describe the mass of the cluster. The final model has an rms of \purple{$0\farcs48$}, and the mass enclosed within 200\,kpc and 500\,kpc is \purple{$\sim1.73\times10^{14}M_{\odot}$ and $\sim5.09\times10^{14}M_{\odot}$}, respectively. We provide a detailed description of the cluster modeling in the following paragraphs. 

\purple{This model is} the first published lens model for MACS0027. The cluster was first reported in \cite{Repp2018} as part of the \hst~SNAPshot surveys (PID: 12166; PI: Ebeling).
We identify four \referee{lensed source galaxies}, of which two have spectroscopic redshifts from MUSE. \referee{Source} \#1 consists of four multiple images arranged in an Einstein-cross configuration, with a spectroscopic redshift measurement of $z_{spec}=3.127$. MUSE also reveals a total of five images with a spectroscopic redshift of $z_{spec}=4.45$. Inspection of the \jwst\ imaging shows that three of these images share the same color and morphology and are spatially arranged in a configuration consistent with expectations from lensing. We label these three images as \referee{Source} \#3. The other two sources at this redshift do not share the morphology or colors of \referee{Source} \#3, but according to the best-fit lens model, they may be counter-images of each other. We thus label these two sources as candidate \referee{Source} \#7, though we do not use it in the lens model. The five MUSE sources are likely associated with two different galaxies at the same redshift, separated by $\sim30$ kpc in the source plane. We discuss these sources further in \autoref{sec:highz}.

\referee{Sources} \#1 and \#2 have a large amount of substructure that is only visible in the \jwst~imaging, and we use these substructure components as additional constraints in the model. We visually identify a radial arc near the BCG, but do not associate it with any \referee{source}, due to the difficulty in securing a reliable spectroscopic redshift for it. We also identify two candidate \referee{lensed source galaxies}, \#5 and \#6, as shown in \autoref{fig.allmodels}. For our final model, we only include \referee{Sources} \#1, \#2, \#3 and \#4, while keeping \referee{Sources} \#5, \#6 and \#7 as candidates. We note that apart from these \referee{lensed source galaxies}, there are several arcs found in \jwst~imaging that have no obvious counter-images at the depth of the available data.

The central peak of the X-ray imaging is generally aligned with the center of the mass peak of the cluster (see \autoref{tab.xray} for the X-ray observation details and \autoref{fig.projmass} for the X-ray contours in cyan and projected mass contours in black). The model optimizes the fits for the redshifts of 2 \referee{lensed source galaxies}. The best-fit optimization of the redshift for \referee{Source} \#2 varies depending on the substructure clump used to perform the fit, where the first clump, \#2.1, has an optimized model redshift of \purple{$5.91_{-0.55}^{+0.01}$}, while the second clump, \#2.2, has an optimized redshift of \purple{$5.17_{-2.37}^{+0.00}$}. The second \referee{source}, \#4, has an optimized redshift of \purple{$4.68^{+0.19}_{-0.35}$}.  

%%%%%%%%%%%%%%%%%%%%%%%%%%%%%%%%%%%%%%%%%%%
%%%%%%%%%%%%%%%%%%%%%%%%%%%%%%%%%%%%%%%%%%%

\subsubsection{RXC\,J2211.7$-$0349} %
RXC\,J2211.7$-$0349 (RXC2211), located at $z=0.397$, is optimized with a total of 10 \referee{lensed source galaxies} (32 multiple images and 45 lensing constraints), where 1 \referee{source galaxy} has a spectroscopic redshift and the remaining 9 redshifts are optimized within the model. A total of 197 cluster member galaxies \purple{ are fitted with a scaling relation based on a reference luminosity $L^*$ for the BCG of the cluster (F814W = 19.20).} The model uses five potentials to describe the mass of the cluster. The final model has an rms of \purple{$0\farcs57$}, and the mass enclosed within 200\,kpc and 500\,kpc is \purple{$\sim2.81\times10^{14}M_{\odot}$ and $\sim7.04\times10^{14}M_{\odot}$}, respectively. We provide a detailed description of the cluster modeling in the following paragraphs.

We update the lens model of RXC2211 that was first presented in \cite{cerny2018}, based on \hst~observations from the ReIonization Lensing Cluster Survey (RELICS; \citealt{coe2019}). The original model consisted of three different \referee{lensed source galaxies}, oriented mainly around the BCG of the cluster. The SLICE \jwst~imaging allows us to add seven new \referee{lensed source galaxies} to this model. The substructure in all 10 \referee{sources} of this model is clearly resolved, which allows us to use the clumps \referee{within} five \referee{source galaxies} as additional constraints in the model. 

\referee{Sources} \#1, \#2, and \#3 correspond to the identifications in \cite{cerny2018}. \referee{Sources} \#4 through \#10 are new identifications that are only visible in the \jwst~imaging. We use the spectroscopic redshift measurement of $z_{spec}=1.501$ from \cite{cerny2018} for \referee{Source} \#1, and allow the model to optimize the redshifts of the other constraints. \purple{All} optimized \referee{sources} have redshifts under $z\sim4$. \referee{Sources} \#9 and \#10, which have an optimized model redshift of $z\sim2$, are located very close to the critical curve, providing a tight constraint on its location. We also identify a candidate lensed galaxy, \referee{Source} \#11, in the south-west part of the cluster, though we do not include it in the final model as its predicted configuration yields additional counter-images that cannot be reliably identified in the \jwst~imaging. \referee{Source} \#4 is a very dusty red galaxy that is bright in the \jwst~imaging. 

The new constraints give definition to the elongated, angularly-warped, projected mass distribution, which is slightly tilted toward the west at both the north and south ends. The complexity of the underlying mass distribution is reflected by the need for two large cluster-scale DM halos at the center of the cluster to accurately reproduce the positions of the \referee{lensed source galaxies and clumps}. An additional two galaxy-scale cluster member halos are also placed near the second- and seventh-brightest galaxies in the cluster core to successfully reproduce the multiple images. Halo 3 adds a significant clump of mass to the west part of the cluster, indicating the presence of possible dark matter substructure. Further improvements to this model rely on the acquisition of more spectroscopic redshifts, especially for the new \referee{lensed source galaxies} identified with~\jwst. 

%%%%%%%%%%%%%%%%%%%%%%%%%%%%%%%%%%%%%%%%%%%
%%%%%%%%%%%%%%%%%%%%%%%%%%%%%%%%%%%%%%%%%%%

\subsubsection{MACS\,J0553.4$-$3342} 
MACS\,J0553.4$-$3342 (MACS0553), located at $z=0.412$, is optimized with a total of 21 \referee{lensed source galaxies} (57 multiple images and 102 lensing constraints), where 4 \referee{source galaxies} have a spectroscopic redshift and the remaining 17 redshifts are optimized within the model. A total of 197 cluster member galaxies \purple{ are fitted with a scaling relation based on a reference luminosity $L^*$ for the BCG of the cluster (F814W = 19.47).} The model uses two potentials to describe the mass of the cluster. The final model has an rms of \purple{$0\farcs66$}, and the mass enclosed within 200\,kpc and 500\,kpc is $\sim1.84\times10^{14}M_{\odot}$ and \purple{$\sim6.15\times10^{14}M_{\odot}$}, respectively. We provide a detailed description of the cluster modeling in the following paragraphs.

We update the lens model that was originally presented in \cite{Ebeling2017} and use the \hst~observations from the ReIonization Lensing Cluster Survey (RELICS; \citealt{coe2019}). The original model consists of 10 different \referee{lensed source galaxies} covering two main subclusters. The SLICE \jwst~imaging allows us to add 11 new \referee{lensed source galaxies} to this model. The substructure in six \referee{sources} of this model is clearly resolved, which allows us to use a total of 102 clumps in the 21 \referee{sources} as constraints in the model. 

\referee{Sources} \#1 through \#10 correspond to the identifications in \cite{Ebeling2017}. \referee{Sources} \#11 through \#21 are new identifications that are only fully confirmed using the \jwst~imaging. All \referee{sources} are comprised of three-imagefolds. We use the spectroscopic redshift measurements of \referee{Sources} \#1 and \#5, and make use of additional VLT/MUSE observations from the Kaleidoscope survey (PID: 0104.A-0801; PI: A. Edge) to confirm Lyman-$\alpha$ emission from \referee{Sources} \#5, \#10 and \#13. We let the model optimize the redshifts of the other \referee{sources}. 

The new constraints confirm the very elongated and strongly bimodal mass distribution of the previous model presented in \citet{Ebeling2017}, and do not require additional complexity. The central peak of the X-ray imaging is aligned with the western peak of the mass distribution, which contains a larger fraction of mass than the eastern portion of the cluster (see \autoref{tab.xray} for the X-ray observation details and \autoref{fig.projmass} for the X-ray contours in cyan and projected mass contours in black). The model optimizes the fits for the redshifts of 17 \referee{sources}. \referee{Source} \#6 has an optimized redshift of \purple{$4.31^{+0.00}_{-0.16}$}, \referee{Source} \#8 has a redshift of \purple{$5.79^{+0.02}_{-0.24}$}, \referee{Source} \#12 has a redshift of \purple{$7.25^{+0.00}_{-0.52}$}, \referee{Source} \#16 has a redshift of \purple{$4.66^{+0.00}_{-0.20}$}, and the remaining \referee{sources} are under $z\sim4$. 

%%%%%%%%%%%%%%%%%%%%%%%%%%%%%%%%%%%%%%%%%%%
%%%%%%%%%%%%%%%%%%%%%%%%%%%%%%%%%%%%%%%%%%%

\subsubsection{MACS\,J0358.8$-$2955} 
MACS\,J0358.8$-$2955 (MACS0358), located at $z=0.425$, is optimized with a total of 6 \referee{lensed source galaxies} (16 images and 16 lensing constraints), where 3 \referee{source galaxies} have a spectroscopic redshift and the remaining 3 redshifts are optimized within the model. A total of 256 cluster member galaxies \purple{ are fitted with a scaling relation based on a reference luminosity $L^*$ for the BCG of the cluster (F814W = 19.32). Only the velocity dispersion for the scaling relation is optimized in the model; the cut radius is fixed to 30 kpc to allow the model to produce a consistent physical result.} The model uses two potentials to describe the mass of the cluster. The final model has an rms of \purple{$0\farcs21$}, and the mass enclosed within 200\,kpc and 500\,kpc is \purple{$\sim2.15\times10^{14}M_{\odot}$ and $\sim5.99\times10^{14}M_{\odot}$}, respectively. We provide a detailed description of the cluster modeling in the following paragraph.

MACS0358, a RELICS \citep{coe2019} cluster, displays a bimodal light distribution elongated in the Nort-East/South-West direction. A BCG is dominating the North-East light clump, whereas the South-West one is dominated by two galaxies with comparable magnitudes. The cluster's 3D geometry and merger history have been investigated by \cite{hsu2013}, suggesting a complex merger of at least three sub-clusters. These authors constructed a mass model using three \referee{lensed source galaxies}. We add three new sources, for a total of six lensed source galaxies, and adopt the spectroscopic redshift measurement \referee{for the} three sources from \cite{hsu2013}. The mass model uses two large-scale DM halos associated with each light concentration. The best-fit mass model has an rms of $0\farcs21$ and our results are in agreement with the former model proposed by \cite{hsu2013}.

The peak of the X-ray imaging is located slightly south of the north-most peak of the mass model (see \autoref{tab.xray} for the X-ray observation details and \autoref{fig.projmass} for the X-ray contours in cyan and projected mass contours in black). The model optimizes the redshifts of three \referee{lensed source galaxies}, all of which are under $z\sim3$. 

%%%%%%%%%%%%%%%%%%%%%%%%%%%%%%%%%%%%%%%%%%%
%%%%%%%%%%%%%%%%%%%%%%%%%%%%%%%%%%%%%%%%%%%

\subsubsection{MACS\,J1621.4$+$3810}\label{photoz} 
MACS\,J1621.4$+$3810 (MACS1621), located at $z=0.4631$, is optimized with a total of 5 \referee{lensed source galaxies} (16 multiple images and 22 lensing constraints). None of the \referee{source galaxies} have a spectroscopic redshift at this time. One \referee{source} (using two clumps) is fixed to its photometric redshift, and the remaining 4 source redshifts are optimized within the model. A total of 150 cluster member galaxies \purple{ are fitted with a scaling relation based on a reference luminosity $L^*$ for the BCG of the cluster (F814W = 19.58).} The model uses three potentials to describe the mass of the cluster. The final model has an rms of \purple{$0\farcs26$}, and the mass enclosed within 200\,kpc and 500\,kpc is \purple{$\sim1.62\times10^{14}M_{\odot}$ and $\sim3.91\times10^{14}M_{\odot}$}, respectively. We provide a detailed description of the cluster modeling in the following paragraphs.

We present a first lens model of the cluster MACS1621, whose discovery was initially reported in \cite{Repp2018}.
We identified five \referee{lensed source galaxies} in this field. \referee{Sources} \#1 and \#2 are resolved in the SLICE \jwst~imaging, with several clumps identified (\autoref{fig.allmodels}). The model uses two clumps as constraints for each of these two \referee{source galaxies}. We could not find any public spectroscopic record of the \referee{lensed source galaxies} in this cluster. We thus rely on a photometric redshift measurement for the brightest image, \referee{Source} \#2, estimated at $z=1.5^{+0.3}_{-0.2}$ using all available \hst~bands (ACS/F606W, ACS/F814W, WFC3IR/F110W, WFC3IR/F140W), to anchor the model and break mass-redshift degeneracies. 

Using the aperture photometric apparent magnitudes (and Gaussian estimated uncertainties) from the arc/clump of interest in available \hst\ and \jwst\ filters, we conduct a Bayesian statistical inference to estimate photometric redshifts. We use the stellar population synthesis modeling package \texttt{Prospector} \citep{2021prospect} to simultaneously infer physical properties of the \referee{lensed galaxy} along with its redshift. 
For our fiducial model, we use a parametric star formation history -- delayed $\tau$ -- model, and fit stellar metallicity, dust attenuation \citep{2000Calzetti}, and total mass formed in the galaxy (with TopHat priors) as nuisance parameters (see \citet{Khullar22} for details on the nominal fitting and modeling prescription used here). 

We also note that the bluest photometry samples the Ly$\alpha$/Ly$\beta$ emission region, and hence we fit an additional free parameter to marginalize the contribution of the \purple{intergalactic medium} (IGM) absorption and Lyman-series flux. The expectation from this process was that Lyman and Balmer breaks in galaxy/clump \purple{spectral energy distributions (SEDs)} would allow us to put limits on redshifts, while simultaneously aiding the lens models. Equally importantly, with limited photometry for these sources, it is critical to simultaneously fit multiple galaxy parameters to respect the age-metallicity-dust degeneracy and the potential for the existence of multiple stellar populations (including emission-line producing stellar regions).

All other source redshifts are set as free parameters in the model.
Although still limited by the lack of spectroscopic redshift constraints, the overall simple geometry of the cluster gives us enough confidence in our ability to accurately describe its shape. This assumption is supported by the overlap of the X-ray imaging peak with the projected surface mass density peak of the cluster (see \autoref{tab.xray} for the X-ray observation details and \autoref{fig.projmass} for the X-ray contours in cyan and projected mass contours in black). The model includes one cluster-scale potential, and the BCG is optimized separately. The BCG potential position is fixed to the center of the BCG, and its shape is free to vary. A third galaxy-scale potential at the cluster redshift is included to account for a foreground galaxy, in the proximity of \referee{Source} \#5. The optimized redshifts for \referee{Sources} \#1-\#4 are all below $z\sim4$.

%%%%%%%%%%%%%%%%%%%%%%%%%%%%%%%%%%%%%%%%%%%
%%%%%%%%%%%%%%%%%%%%%%%%%%%%%%%%%%%%%%%%%%%

\subsubsection{RCS2\,0327$-$1326} \label{sec.RCS0327} 
We present two lens models for RCS2\,0327$-$1326 (RCS0327), located at $z=0.564$, using two different algorithms: \lenstool\ and \wslap. The \lenstool\ model is optimized with a total of 23 \referee{lensed source galaxies} (61 multiple images and 184 strong lensing constraints), where 6 \referee{source galaxies} have spectroscopic redshifts, and the remaining 17 are optimized within the model. The \wslap\ model is optimized using only the 6 \referee{sources} with a spectroscopic redshift (16 multiple images and 137 strong lensing constraints). Both models are constructed using the same catalogue of 132 cluster member galaxies.

The field hosts a bright, highly magnified, $\sim38$\arcsec\ giant arc of a $z=1.701$ galaxy that has been studied extensively \citep{Sharon12, wuyts14, Bordoloi16, Lopez18}. 
The giant arc in the North of the field is composed of three partial images that merge at two critical curve crossings, and has a full counter-image in the South. A fifth demagnified image was detected by \cite{Sharon12} $0\farcs6$ North of the BCG. 

\paragraph{\lenstool\ model}
We improve upon the \hst+MUSE-based model that was used in \cite{Lopez18} and later work. That model is based on and improves the \cite{Sharon12} model by adding constraints from two lensed sources and three spectroscopic redshifts from MUSE to the previous model of \cite{Sharon12}. The \lenstool\ model is optimized in the source plane, due to the large number of constraints.
The SLICE \jwst~imaging reveals or confirms 19 new lensed sources, including six new radial arcs to add to the one formerly known \citep{Sharon12}. In addition, we identify and map 42 clumps in the multiple images of the giant arc. We adopt MUSE spectroscopic redshifts from \cite{Lopez18} for \referee{Source} \#7 (S7a/S7b in \citealt{Sharon12}) at  $z_{spec} = 2.82624$; \referee{Source} \#4 at  $z_{spec} = 2.73$; and \referee{Source} \#5 at $z_{spec} = 5.2$. 
We discover a new radial arc in the MUSE data, labeled as \referee{Source} \#6, with a bright Ly$-\alpha$ emission line and CIII placing this arc at $z_{zpec}=3.518$. A faint counterpart appears in the \jwst\ imaging at the same location; the counter-image is predicted to be outside of the MUSE field of view.
We use all the \referee{lensed source galaxies and clumps} in \autoref{tab.MultipleImages} as constraints in the lens model. More arc candidates are visible in the data, and we leave their confirmation and possible use for further refining of the lens model to future work.

The \lenstool\ model optimizes seven DM halos, including two DM cluster-scale halos at the cluster core, one of which is constrained to within 2\arcsec\ of the BCG; three galaxy-scale halos-- two cluster members within the West end of the giant arc, and one South of the giant arc -- are allowed to be solved for separately from the scaling relations, as they are likely contributing a local shear effect on either end of the giant arc. Another small halo is positioned at the location of a small group of galaxies. Finally, a halo in the East is necessary in order to reproduce the faint arc (A in \citealt{Sharon12}). \purple{The 132 cluster member galaxies \purple{ are fitted with a scaling relation based on a reference luminosity $L^*$ for the BCG of the cluster (F814W = 20.17).} The best-fit rms is \purple{$0\farcs77$}. Due to the large number of constraints along the giant arc, and the equal weights assigned to all the constraints, the model is dominated by the constraints of the giant arc and is thus more tuned to producing accurate results in its vicinity, which could be at the expense of other regions in the field of view. The mass enclosed within 200\,kpc and 500\,kpc is $\sim1.91\times10^{14}M_{\odot}$ and $\sim5.14\times10^{14}M_{\odot}$}, respectively.

\paragraph{\wslap\ model}
The large number of \referee{clumps} available for \referee{Source} \#1 enables the use of free-form methods such as \wslap, which require a large number of constraints to build functional models (see Section~\ref{sec:wslap}). The \wslap\ model is derived from the lensed galaxies in the \lenstool\ model described above that have a spectroscopic redshift; any constraints in the \lenstool\ model that do not have this measurement are not included in the \wslap\ model. The model contains $536+1+2\times51$ free parameters. Out of these, 536 of them correspond to the Gaussian functions that are distributed on a multi-resolution grid (derived iteratively starting from a regular grid of $20\times20$ Gaussians). One free parameter accounts for a renormalization factor to the mass of the member galaxies. The mass of these galaxies traces the observed light in the F150W2 filter and it is assumed as a fiducial mass that is later optimized by \wslap. Finally, the last $2\times51$ free parameters correspond to the $x$ and $y$ positions in the source plane of the \referee{clumps} in \referee{Source} \#1 and the other lensed galaxies used as constraints. The final solution is obtained after optimizing 200,000 steps ($\approx 5$ minutes on a laptop), once convergence is achieved.  The model contains $\sim2.02\times10^{14}$\,\Msun within 200\,kpc from the center and $\sim5.34\times10^{14}M_\odot$ within 500\,kpc from the center. 

The \lenstool\ and \wslap\ models, which are constructed using a subset of the same constraints, result in similar overall mass distributions, particularly within the strong lensing region confined by the tangential critical curve. The difference between the models for both enclosed mass measurements is less than $\sim4\%$. Nevertheless, minor differences between their resulting critical curves are observed (\autoref{fig.allmodels}). The discrepancy is smaller in regions that are well-constrained in both models, e.g., the critical curve crossing of the giant arc of \referee{Source} \#1, which has numerous clumps, and near other arcs with spectroscopic redshift. Notably, while the East and West regions are relatively well constrained in the \lenstool\ model, the lack of spectroscopic redshifts leaves these regions relatively under-constrained for \wslap. A deep comparison between these two algorithms is beyond the scope of this paper and will be left to future work.

%%%%%%%%%%%%%%%%%%%%%%%%%%%%%%%%%%%%%%%%%%%
%%%%%%%%%%%%%%%%%%%%%%%%%%%%%%%%%%%%%%%%%%%

\subsubsection{MACS\,J2129.4$-$0741} 
MACS\,J2129.4$-$0741 (MACS2129), located at $z=0.589$, is optimized with a total of 17 \referee{lensed source galaxies} (53 multiple images and 53 lensing constraints), where 13 \referee{source galaxies} have a spectroscopic redshift confirmation \citep[][]{zitrin2015, Caminha2019,jauzac2021} and the remaining 4 \referee{source} redshifts are optimized within the model. 
The mass model includes two cluster-scale potentials to describe the overall mass distribution of the cluster.
A total of 293 cluster member galaxies are fitted with a scaling relation optimizing two parameters.  \purple{The reference luminosity $L^*$ used in this relation is associated with the BCG of the cluster (F814W = 17.94).} Five cluster galaxies are also modeled independently due to their proximity to multiple images \citep[all details can be found in][]{jauzac2021}. Newly identified multiple images do not require the independent modeling of other cluster members.
Our mass model is optimized using the \Lenstool\ software \citep{jullo07}. \purple{The best-fit rms is $0\farcs79$.}
The masses enclosed within 200\,kpc and 500\,kpc are $\sim 1.84\times10^{14}M_{\odot}$ and $\sim5.32\times10^{14}M_{\odot}$, respectively. We provide a detailed description of the cluster modeling in the following paragraph.

The model presented in this work is based on the model presented in \cite{jauzac2021} and includes three newly identified \referee{lensed source galaxies} from the SLICE-\jwst~imaging (\referee{Sources} \#16, \#17 and \#18). 
This elongated cluster presents one dominant elliptical cluster-scale potential. In addition, we relax the BCG potential and four other galaxy potentials near \referee{lensed source galaxies} in this central region. One more group-scale potential in the North-West region of the cluster is added, matching the light distribution of an overdensity of bright cluster galaxies with diffuse emission. This component is interpreted as an infalling group, even if the X-ray imaging does not show a gaseous counterpart at the position of the galaxy group (see \autoref{tab.xray} for the X-ray observation details, and \autoref{fig.projmass} for the X-ray contours in cyan and projected mass contours in black). Furthermore, we note that the gas peak in the X-ray distribution coincides with the main lensing mass peak, which corresponds to the location of the BCG. 
This updated mass model is in excellent agreement with the model it is based on,
with the rms of the SLICE best-fit mass model at 0.79\arcsec, in comparison to 0.80\arcsec\ for the \cite{jauzac2021} best-fit model rms.  
The final model optimizes the redshifts of \referee{Sources} \#5, \#16, \#17 and \#18. 
\referee{Source} \#18 has a best-fit optimized redshift of 5.53$^{+0.47}_{-0.19}$. The other \referee{sources} have optimized redshifts below 1.7.

%%%%%%%%%%%%%%%%%%%%%%%%%%%%%%%%%%%%%%%%%%%
%%%%%%%%%%%%%%%%%%%%%%%%%%%%%%%%%%%%%%%%%%%

\subsubsection{PSZ1\,G091.83$+$26.11} \label{sec.PSZG091}  
%%%%%%%%%%%%%%%%%%%%%%%%%%%%%%%%%%%%%%%%%%%

We construct \purple{a} preliminary lens model of PSZ1\,G091.83$+$26.11 (PSZ1G09), located at $z=0.822$\purple{, using both \lenstool\ and \wslap. The \lenstool\ model is optimized with a total of 5 \referee{lensed source galaxies} (19 multiple images and 89 strong lensing constraints). The \wslap\ model is optimized with a total of two \referee{sources} (9 multiple images and 71 strong lensing constraints). An additional 7 candidate \referee{lensed source galaxies} or galaxy-galaxy lensing \referee{instances} are identified within the field of view, but are not used in either model. There are no spectroscopic redshifts available for lensed galaxies in this field at the time of writing. Both models are constructed using the same 217 cluster member galaxies. }

This is one of the most massive clusters known at $z>0.8$ \citep{PSZ2} and hence deserves special attention. Clusters at this redshift are in an active phase of formation and often in a non-relaxed state. 
PSZ1G09 shows a bimodal structure both in SZ maps \citep{PSZ2,Artis2022} as well as in the X-rays \citep{Artis2022}, thus suggesting an active merger. This interpretation is supported by radio observations that show prominent radio shock emission on the East side of the cluster \citep{DiGenaro2023}. To date, no lens model is available in the literature. The \jwst\ observations show a very large galaxy imaged into 4 counter-images (\referee{Source} \#1). This galaxy has two components: a blue and compact nucleus that is already clearly visible in earlier observations with \hst, and a very red extended tail, which is barely detectable in previous \hst\ observations but is very prominent in F322W2 and possibly harbours copious amounts of dust. These types of blue-red pairs have been observed multiple times with \jwst, often corresponding to interacting galaxies at $z>3$ \citep{diego23}. 
Combining \hst\ and \jwst\ photometry, we derive photometric redshifts (as outlined in \autoref{photoz}) for all 4 images of \referee{Source} \#1 finding the redshift of the two most reliable images to be 4.39 and 4.44. We adopt a redshift of $z_{phot}=4.4$ for this \referee{source}. In the new \jwst\ data, we identify 10 additional \referee{lensed source} candidates. For \referee{Source} \#2,  composed of two radial arcs and a tangential arc, we derive a photometric redshift of $z_{phot}=2.8$. All the remaining \referee{sources} lack reliable photometric redshift estimates since they are too faint to be detected in \hst\ observations.
Only the \referee{sources} with redshift estimates can be used as constraints in the \wslap\ algorithm. 

To provide sufficient constraining power in the East, the \wslap\ model assumes a redshift $z\sim3$ for a third \referee{source}, \referee{Source} \#8. The \lenstool\ model can use \referee{sources} without a redshift estimate as constraints by including the redshift as a free parameter to be solved by the model. The \lenstool\ model uses the secure arcs as constraints. We describe the two models below. 

\paragraph{\wslap\ model}
We model PSZ1G09 with the hybrid algorithm \wslap. We divide the member galaxies into two layers, with the BCG in Layer 1 and all the remaining members in Layer 2. The smooth component is modeled by a combination of Gaussian functions. A first model is derived on a grid of $20\times20$ points, with a Gaussian centered on each grid point. This solution is used to construct a new grid for the Gaussian centers where more Gaussians (with smaller full-width at half maximums (FWHMs)) are placed in the regions with more mass. The new grid lowers the number of Gaussians (from 400 to 267), and increases the resolution near the regions of higher density. The number of free parameters in the model is then $267+2+2\times11=291$ (267 Gaussians, 2 layers, and the positions in the source plane of 11 lensed galaxies). The bimodal structure of this cluster is evident in both the X-ray imaging and the projected surface mass density map. The X-ray peak is located around the middle of the two mass peaks, lending further evidence to the cluster's status as an active merger (see \autoref{tab.xray} for the X-ray observation details and \autoref{fig.projmass} for the X-ray contours in cyan and projected mass contours in black).
The model has an rms of 1\farcs06. The mass enclosed within 200\,kpc and 500\,kpc is $\sim1.80\times10^{14}M_{\odot}$ and $\sim6.30\times10^{14}M_{\odot}$, respectively.

We compute a second \wslap\ model to explore a secondary mass distribution in the East, using the candidate \referee{Source} \#8 as constraint with an assumed redshift of $z\sim3$. From this model, the mass enclosed within 200\,kpc and 500\,kpc is $\sim 1.84\times10^{14}M_{\odot}$ and $\sim 7.05\times10^{14}M_{\odot}$, respectively.

\paragraph{\lenstool\ model}

The \lenstool\ model of PSZ1G09 uses the same constraints as the \wslap\ model for \referee{Sources} \#1 and \#2. Since \lenstool\ can include \referee{sources} without photometric redshifts as constraints with free redshift parameters, we are able to use \referee{Source} \#3 to constrain the cluster center, and \referee{Sources} \#4 and \#5 to constrain the north part of the cluster core. We used an iterative modeling approach, starting with a lens plane composed of red-sequence-selected cluster member galaxies and a single cluster-scale halo with broad priors. As expected from the arc geometry of the images of \referee{Source} \#1, a single halo model resulted in a poor solution. We iteratively added cluster-scale halos, increasing the model complexity, until no significant improvement was recorded. The best-fit model converged on three cluster-scale halos: two near the main strong lensing core, and one east of the main strong lensing core. The slope and normalization parameters of the east halo were allowed to vary, but its position was fixed to the location of the brightest cluster galaxy, $\sim50''$ east of the main strong lensing core, at R.A., Decl. = [277.816360, 62.245004], and we assumed a circular mass distribution. The arc candidates near this halo were not used as constraints in the \lenstool\ model. We additionally freed some of the parameters of the two galaxies near the radial arcs of \referee{Source} \#2, to increase the model flexibility at the center (see \autoref{tab.bestfitparams}). \purple{The 218 cluster member galaxies \purple{ are fitted with a scaling relation based on a reference luminosity $L^*$ for the BCG of the cluster (F110W = 20.25).} The best-fit model has an image-plane rms of \purple{$0\farcs47$}. The mass enclosed within 200\,kpc and 500\,kpc is \purple{$\sim1.89\times10^{14}M_{\odot}$ and $\sim7.61\times10^{14}M_{\odot}$}, respectively. It uses 5 \referee{sources} as strong lensing constraints: \referee{Source} \# 1 at $z_{phot}=4.4$ with four images, each with 18 clumps; \referee{Source} \#2 at $z_{phot}=2.8$ with five images; \referee{Source} \#4 (2 clumps, \purple{ best-fit optimized redshift below $z\sim4$}) and \#5 (3 clumps, best-fit optimized redshift of $4.66_{-0.17}^{+0.03}$}) with three images each; and Source \#3, which has three secure images (best-fit optimized redshift of $4.47^{+0.00}_{-0.17}$). The fourth is left as a candidate and not used as constraint since there is more than one possible arc near the location predicted by the lens model from the other three images. Acquiring a reliable redshift measurement (i.e. a spectroscopic measurement) for this source would likely allow the model to identify the correct counter-image from the three possible candidate images. 

\autoref{fig.allmodels} shows the critical curves of the \wslap\ model (white) and \lenstool\ model (black) for a source at $z=4.4$. Although using very different approaches, the two models yield similar critical curves, with the main discrepancies occurring in regions with less constraints. An examination of the mass measurements between the two modeling methods shows that the \wslap\ model created with the same constraints as \lenstool\ measures a mass that is $\sim5\%$ lower than the \lenstool\ model within 200 kpc and $\sim19\%$ lower within 500 kpc. This difference originates from the lack of reliable constraints in the Eastern half of the cluster. The \lenstool\ model chooses to place a parametric halo fixed to the BCG of this portion of the cluster, while the \wslap\ model created with the same constraints as \lenstool\ cannot constrain a mass halo in this region, and thus does not include one. The second \wslap\ model, which is created with different constraints than the \lenstool\ model, elects to place a free-form mass halo around the location of candidate \referee{Sources} \#8 and \#9. The agreement between this second model and the \lenstool\ model at 500 kpc is better, with only a $\sim9\%$ difference between the two models. This comparison highlights the importance of the Eastern mass component of this cluster in constraining the model. The acquisition of at least one spectroscopic redshift will enable both modeling methods to explore the parameter space in this region more rigorously. 
%
%%%%%%%%%%%%%%%%%%%%%%%%%%%%%%%%%%%%%%%%%%%

\subsubsection{RX\,J0152.7$-$1357} 
RX\,J0152.7$-$1357 (RX0152), located at $z=0.8269$, is optimized with a total of 7 \referee{lensed source galaxies} (20 multiple images and 26 lensing constraints), where 1 \referee{source galaxy} has a spectroscopic redshift and the remaining 8 \referee{source} redshifts are optimized within the model. A total of 131 cluster member galaxies \purple{ are fitted with a scaling relation based on a reference luminosity $L^*$ for the BCG of the cluster (F160W = 20.64).} The model uses one cluster-scale potential to describe the mass of the cluster. The final model has an rms of $0\farcs45$, and the mass enclosed within 200\,kpc and 500\,kpc is $\sim1.14\times10^{14}M_{\odot}$ and $\sim3.44\times10^{14}M_{\odot}$, respectively. We provide a detailed description of the cluster modeling in the following paragraphs.

In this work, we present a new mass model of RX0152 based on the SLICE \jwst~imaging. This cluster presents three overdensities of galaxies, highlighting a dominant cluster-scale halo and two group-scale halos in the South-West and West of the cluster. These three structures are apparent in the X-ray emission as highlighted by the contours of the X-ray surface brightness in \autoref{fig.projmass}. These structures were already reported by a previous lensing analysis presented in \citet{Acebron2019}.

Thanks to the new \jwst~imaging, we identify one new \referee{lensed source galaxy} located near a cluster member overdensity,  South-West of the cluster. The \referee{images of this source are} barely visible in the reddest band of the RELICS imaging (i.e. WFC3/F160W), but it appears clearly in the \jwst/F322W2 band as a galaxy-scale lensing feature where only a part of the galaxy disk is multiply imaged. The color and the apparent size of the singly lensed body of the galaxy likely highlight a dusty galaxy at $z\gtrsim2$. We confirm morphologically the six multiply-imaged galaxies used as constraints in \citet{Acebron2019}. 
We could secure more accurate positions of the multiple images thanks to star-forming regions clearly appearing in the \jwst~imaging. Two of the lensed galaxies show more than one clump; in total we therefore have nine \referee{lensed source galaxies} as lensing constraints for our mass model.

Our \lenstool~model includes $131$ halos to account for the cluster members that we identified with a red sequence fitting based on the \hst\ imaging in the ACS/F625W and WFC3/F160W bands. 
Based on this selection, we measure their position, position angle and ellipticity based on the \jwst/F322W2 band. In addition, we add one cluster-scale halo to represent the smooth DM component of the main cluster. We try to add another halo in the South-West to account for the galaxy-group mentioned earlier, but the lack of \referee{lensing constraints} in this regions leads to an unrealistic mass of this component. Unfortunately, only half of the structure is visible in \jwst~imaging, limiting our ability to identify new multiply imaged candidates.

Our model is similar to the \lenstool~model presented in \citet{Acebron2019} in its parameterization, one cluster-scale DM halo and the cluster galaxy-scale components. 
However, we are not modeling any galaxy outside of the cluster member relation. Indeed, in \citet{Acebron2019}, they model one cluster galaxy individually due to its proximity to \referee{the images of} \referee{Source} \#1.
Using the same \referee{sources} as \citet{Acebron2019}, our best-fit model has an RMS of 0.47\arcsec, a slightly lower value than the rms of 0.52\arcsec\ reported by \cite{Acebron2019}.
When including the newly identified \referee{lensed source galaxy} presented earlier, our best-fit mass model has an rms of 0.45\arcsec. \purple{\referee{Source} \#6 has a best-fit optimized redshift of $4.39_{-0.46}^{+0.43}$; \referee{Source} \#7 is optimized to $4.35^{+0.44}_{-0.47}$; \referee{Source} \#11 is optimized to $4.45^{+0.00}_{-2.58}$; and all other optimized redshifts are below $z\sim3.$}

%%%%%%%%%%%%%%%%%%%%%%%%%%%%%%%%%%%%%%%%%%%
%%%%%%%%%%%%%%%%%%%%%%%%%%%%%%%%%%%%%%%%%%%

\subsubsection{SPT-CL\,J0516$-$5755} 
SPT-CL\,J0516$-$5755 (SPT0516), located at $z=0.9656$, is optimized with a total of 10 \referee{lensed source galaxies} (25 multiple images and 53 lensing constraints), where the model is anchored using a photometric redshift estimate for 1 \referee{source galaxy} and the remaining 9 \referee{sources} are optimized within the model. A total of 247 cluster member galaxies \purple{ are fitted with a scaling relation based on a reference luminosity $L^*$ for the BCG of the cluster (F150W2 = 19.30).} The model uses three potentials to describe the mass of the cluster. The final model has an rms of \purple{$0\farcs21$}, and the mass enclosed within 200\,kpc and 500\,kpc is \purple{$\sim1.74\times10^{14}M_{\odot}$ and $\sim5.10\times10^{14}M_{\odot}$}, respectively. We provide a detailed description of the cluster modeling in the following paragraphs.

We present the first SL model of SPT0516. Archival \hst\ data exist only in F606W for this field. We select cluster member galaxies by their color in ACS/F606W-NIRCam/F150W2, and magnitudes as measured from the NIRCam/F150W2 band. 
We identify 10 lensed sources with high confidence to be used as lensing constraints. The elongated projected mass distribution results in naked cusp lensing configurations for virtually all the \referee{lensed sources}, with three images of each source. We only identify a cental fourth image for \referee{Source} \#1, near a cluster member at the cluster core. 
The most prominent \referee{source} in the SLICE \jwst~imaging is entirely absent from the \hst/F606W data -- a red dusty galaxy with  distinct morphology and obvious dust lane. Interestingly, the Eastern image of this galaxy features an emission clump that is missing from the other images. We discuss this candidate transient in \autoref{sec.sn}. 

Unfortunately, no spectroscopic redshifts were measured for lensed galaxies in this field. 
An attempt to obtain spectroscopy in this field in 2024 October (soon after the \jwst\ observations, and prompted by the discovery of the candidate transient, see \autoref{sec.sn}) failed due to weather conditions. We were awarded time on the Magellan telescope in 2025A (U. of Michigan allocation, PI: Sharon) to obtain spectroscopic redshifts of multiple images in this field, and will revisit the lens model once spectroscopic redshifts are secured.   

\begin{figure*}
\centering
\begin{minipage}{0.2\linewidth}
    \includegraphics[width=1\linewidth]{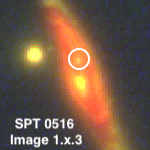}
\end{minipage}
\begin{minipage}{0.65\linewidth}
    \includegraphics[width=0.98\linewidth]{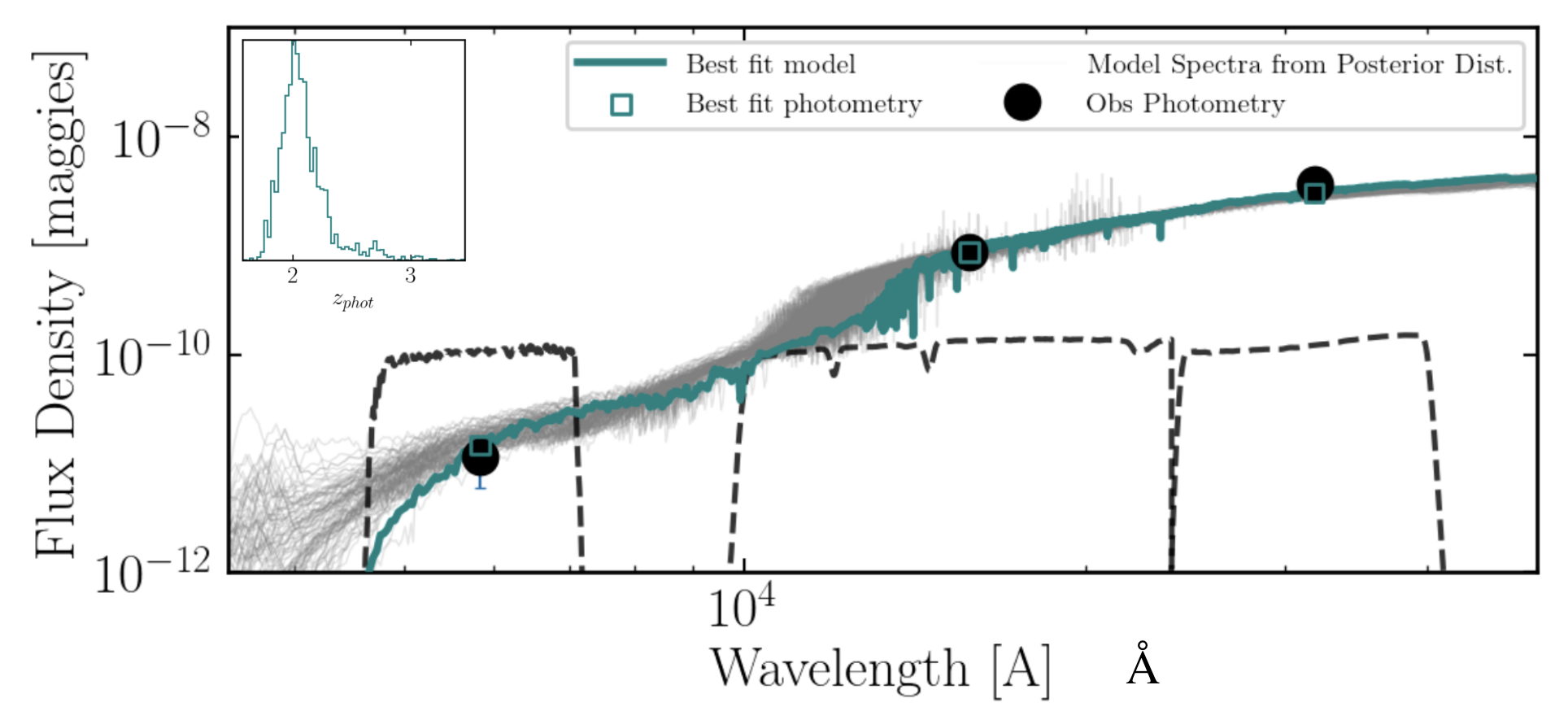}
\end{minipage}
\begin{minipage}{0.2\linewidth}
    \includegraphics[width=1\linewidth]{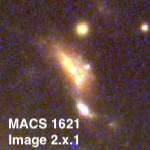}
\end{minipage}
\begin{minipage}{0.65\linewidth}
    \includegraphics[width=1\linewidth]{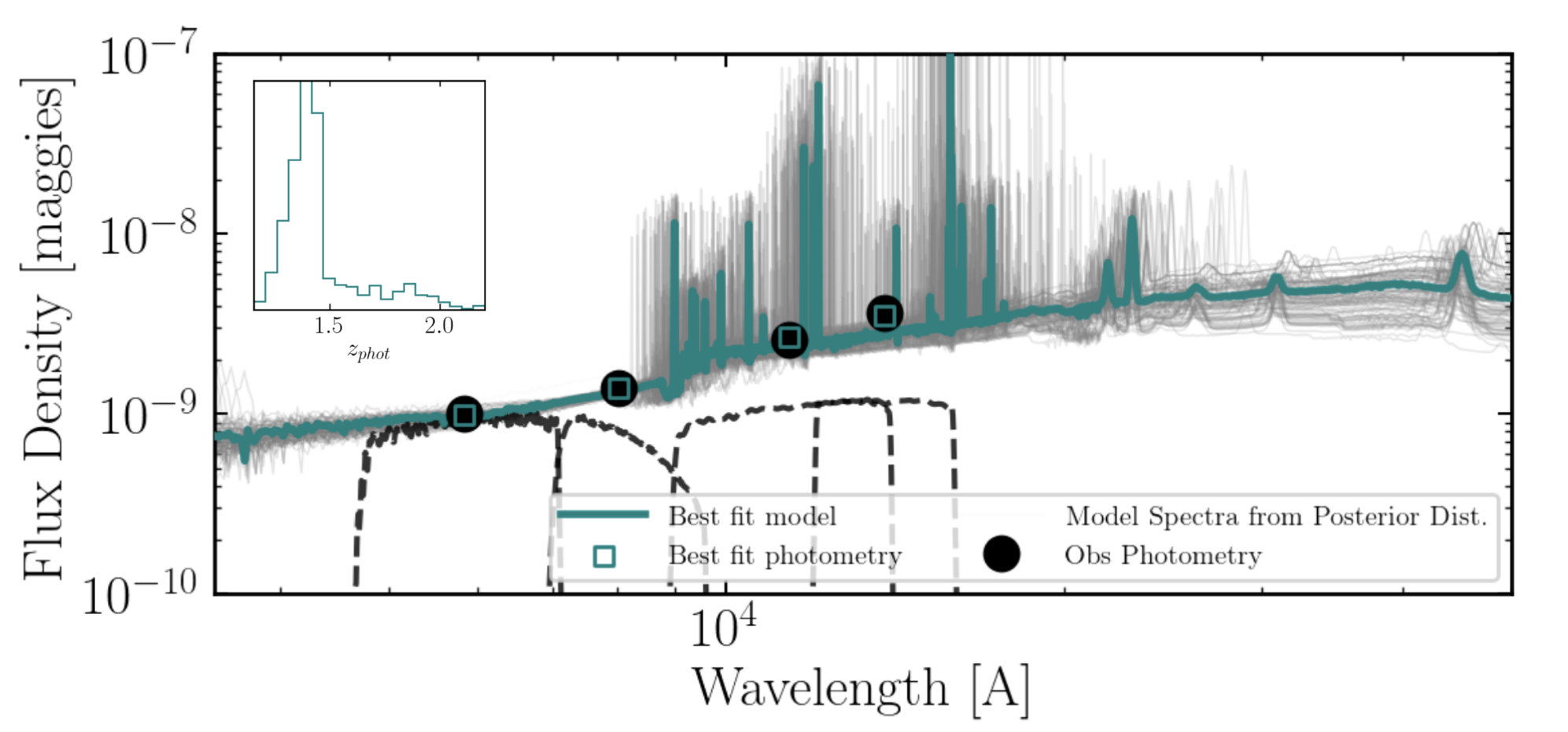}
\end{minipage}
    \caption{\textit{Top:} SED analysis of Image \#10.1 in SPT-CL\,J0516$-$5755, shown in the postage stamp on the left. The candidate transient is circled in white. The photometry is measured in %\hst/F606W, \jwst/F150W2,
    ACS/F606W, NIRCam/F150W2, and NIRCam/F322W2. The best-fit redshift is consistent with $z_{phot}=2.0^{+0.7}_{ -0.4}$, as can be seen in the inset. \textit{Bottom:} SED analysis of \referee{Source} \#2 in MACS\,J1621.4$+$3810, shown in the postage stamp on the left. The photometry is measured in ACS/F606W, ACS/F814W, WFC3IR/F110W, and WFC3IR/F140W. The estimated redshift measurement is set at $z_{phot}=1.5^{+0.3}_{-0.2}$.  }
    \label{fig:sed}
\end{figure*}

To compute the model presented here, we anchor the model at the photometric redshift of \referee{Source} \#1 and leave the redshifts of all the other \referee{sources} as free parameters in the model. We then test the results of the model against the photometric redshifts of other \referee{sources}. 
Photometric redshifts were estimated using \texttt{Prospector} (see \autoref{photoz}) to fit the photometry of Image \#10.1, F606W$=27.36\pm0.44$ (consistent with no detection), F150W2$=22.65\pm0.02$, and F322W2$=21.08\pm 0.01$. \autoref{fig:sed} shows the fit results. The redshift posterior distribution of \referee{Source} \#1 is consistent with $z_{phot}=2.0^{+0.7}_{-0.4}$. \referee{Sources} \#2 and \#3 have broader posterior distributions, favoring redshifts $2\lesssim z \lesssim 4$. We produce two models, one assuming that \referee{Source} \#1 is at $z=2.0$, and one assuming $z=2.5$. Both models produce satisfactory results in terms of reproducing the lensing observables and low rms. However, the model that assumes $z_{src1}=2.0$ predicts redshifts for \referee{Sources} \#2 and \#3 that are inconsistent with the photo-z estimates. We therefore adopt the $z_{src1}=2.5$ model moving forward, and report the model parameters and lensing outputs from this model in \autoref{tab.bestfitparams}.

%\begin{figure}[t!]

The SLICE \jwst~and single-band \hst\ imaging allow us to identify and map substructure in more than half of the \referee{sources}, multiplexing the number of lens model constraints.  
As shown in \autoref{fig.projmass}, the overall shape of the projected mass distribution is elongated in the North-West/South-East direction. Our final best-fit mass model has a rms of \purple{$0\farcs21$}. \referee{Source} \#4 has an optimized redshift of \purple{$5.32^{+0.08}_{-0.41}$} for clump 1 and \purple{$5.51^{+0.00}_{-0.86}$} for clump 2. The rest of the optimized redshifts for the \referee{sources} in this cluster are below $z\sim4$.

\subsubsection{Candidate Transient in SPT-CL\,J0516$-$5755}
\label{sec.sn}
\begin{figure}[h]
\begin{minipage}{1\linewidth}
    \includegraphics[width=1\linewidth]{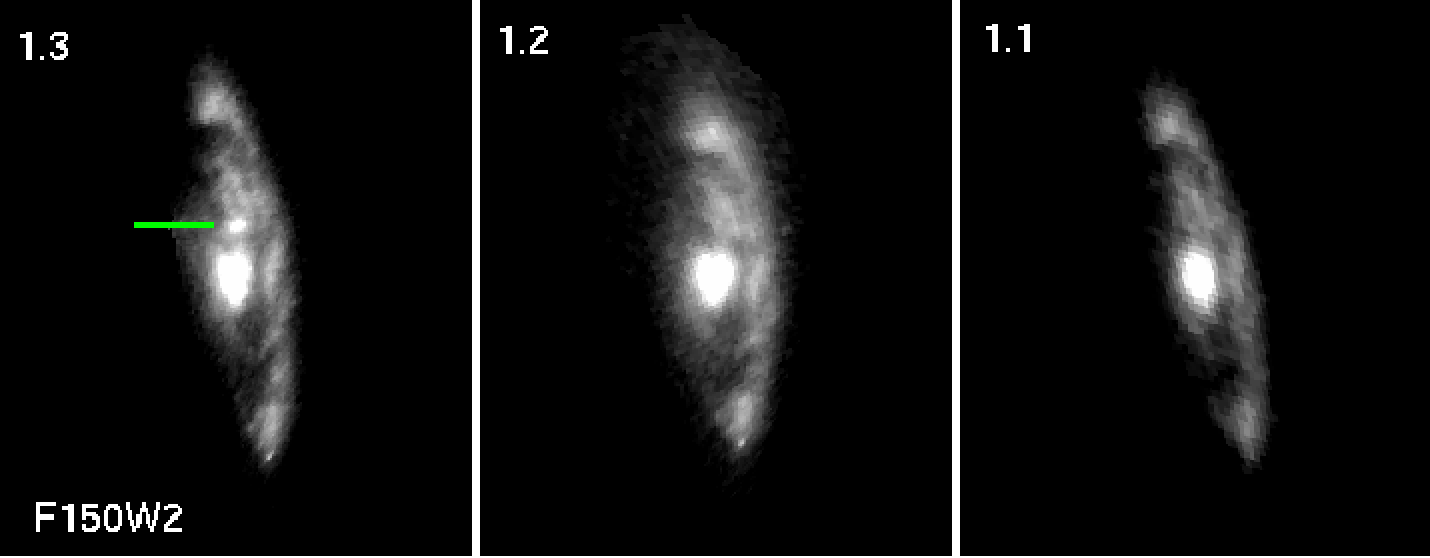}
\end{minipage}
    \caption{Source projection of the three multiple images of \referee{Source} \#1 in %SPT-CL\,J0516$-$5755
    SPT0516, produced by ray-tracing the NIRCam/F150W2 image through the best-fit lens model. A green line in the left panel points to the candidate transient, which appears in Image \#1.3 of the source, and missing from the two counter-images. The host galaxy is assumed to be at $z_{phot}=2.5$. Spectroscopic confirmation is pending. The host galaxy is not visible in the %\hst
    ACS/F606W band, likely due to obscuration by dust. The model predicts that, if indeed a transient, it will reappear in Image \#1.2 $\sim530$ days after its first appearance in Image \#1.3.}
    \label{fig:0516zphot}
\end{figure}

The most prominent lensed \referee{source galaxy} in this field is \referee{Source} \#1, lensed into a set of three full images of a red dusty galaxy appearing South of the BCG, and a fourth partial image clearly seen in the light of a cluster member galaxy $1\farcs8$ North-East of the BCG. We show the reconstructed source projection of the three full images in~\autoref{fig:0516zphot}. Interestingly, in Image \#1.3, we identify a point-like emission clump that does not appear in any of the other images. Owing to the low distortion, good resolution, and clear symmetry between the three full images, the identification is unambiguous. The candidate could possibly be a lensed supernova at $z\sim2$, adopting the photometric redshift of its host galaxy. 
The lens model predicts the time delay between the multiple images of the lensed source. If this is indeed a transient and not a foreground interloper, the model predicts that this is the second image to appear; the first one occurred in Image \#1.1 (the Western image) $\sim4000$ days prior, and the next appearance will be in Image \#1.2 in $560$ days. The model is limited by the lack of spectroscopic redshifts, however, estimating the time delays with a model that anchors this source at $z=2$ instead of $z=2.5$ yields similar results, predicting that the first image occurred $\sim4300$ days before the second (current) one, and the third image will appear $530$ days after the second image. We therefore predict that a counter-image of the candidate transient event will be visible in Image \#1.2 of the galaxy in the next two years.

%%%%%%%%%%%%%%%%%%%%%%%%%%%%%%%%%%%%%%%%%%
%%%%%%%%%%%%%%%%%%%%%%%%%%%%%%%%%%%%%%%%%%%

\subsubsection{SPT-CL\,J2011$-$5228} 
SPT-CL\,J2011$-$5228 (SPT2011), located at $z=1.064$, is optimized with a total of 3 \referee{lensed source galaxies} (12 multiple images and 96 lensing constraints), where 1 \referee{source galaxy} has a spectroscopic redshift and the remaining 2 \referee{sources} are optimized within the model. A total of 112 cluster member galaxies \purple{ are fitted with a scaling relation based on a reference luminosity $L^*$ for the BCG of the cluster (F125W = 20.63).} The model uses five potentials to describe the mass of the cluster. The final model has an rms of \purple{$0\farcs33$}, and the mass enclosed within 200\,kpc and 500\,kpc is \purple{$\sim2.51\times10^{14}M_{\odot}$ and $\sim7.73\times10^{14}M_{\odot}$}, respectively. We provide a detailed description of the cluster modeling in the following paragraphs.

The SLICE \jwst~imaging for this cluster reveals an impressive amount of substructure in the main lensed \referee{source galaxy}. These clumps appear across five main multiple images, including a radial arc. The resolution of \jwst~reveals this radial arc in precise detail, showing that it is an almost-exactly mirrored counterpart of the southernmost arc. \jwst\ also demonstrates that the South-West image is actually a merging pair that is intersected by the critical curve of the model, with a multiplicity of up to three within this image alone, depending on the location of the substructure clumps in the lensed galaxy. To construct a lens model for this cluster, we use the spectroscopic redshift measurement for the main \referee{lensed source galaxy} from \cite{collett2017}. We reproduce the lens model from this paper and add 25 different clumps to the model. These clumps are only clearly resolved in \jwst. To reproduce the positions of each of the clumps, a total of two cluster-scale DM halos are required in the \lenstool~model, with one halo centered on the BCG and the second one located West of the cluster, near the merging pair. This extra halo also partially accounts for the shear effect produced by several large foreground galaxies, which was previously noted in \cite{collett2017}. A third DM halo is also added near the foreground galaxy by image \#1.x.3 to account for the shearing effect of this galaxy on this image. We also optimize the second brightest cluster member galaxy and the galaxy near the second image of the main \referee{source} (\referee{Source} \#1), as these galaxies are likely contributing to the mass producing the lensing effect. \referee{Source} \#2 is spatially associated with \referee{Source} \#1 but has a different multiplicity than the rest of the substructure clumps within the lensed galaxy. We assign its redshift to be the same as the spec-z for \referee{Source} \#1 to reproduce the observed lensing constraints. \referee{Source} \#3 has an optimized model redshift below $z\sim4$.

%%%%%%%%%%%%%%%%%%%%%%%%%%%%%%%%%%%%%%%%%%%

%%%%%%%%%%%%%%%%%%%%%%%%%%%%%%%%%%%%%%%%%%%
\section{Results} \label{sec:results}

The clusters presented in this paper represent the first effort in cluster mass modeling from the first few months of observations of the SLICE program, and were not deliberately selected to be a representative subsample. Nevertheless, they span both a wide range of redshifts (from $z\sim0.2$ to $z\sim1$) and masses ($M_{500}\sim2 \times 10^{14}M_{\odot}$ to $M_{500}\sim12 \times 10^{14}M_{\odot}$), covering nearly the entire range of the parent sample. The lensing properties of the subsample are thus worth considering in aggregate. In this section, we discuss several of the key results from the SL models, including the projected surface mass density measurements, the alignment of the mass peaks with the \textit{Chandra} X-ray observations, the identification of new constraints facilitated by inspection of the \jwst~images, and an estimate of the lensing strength of each cluster. 

%%%%%%%%%%%%%%%%%%%%%%%%%%%%%%%%%%%%

\subsection{Mass}

We compute the projected mass within 200\,kpc and 500\,kpc for each strong lensing cluster, which we report in \autoref{tab.clustermass}. We use the mass enclosed within these distances as our primary metric since it is difficult to use strong lensing models to approximate quantities like $M_{500}$ without drawing potentially non-physical assumptions about the shape of the overall cluster mass distribution. SL only constrains the central cluster mass. 
This limitation can be seen in the projected mass surface density plots shown in \autoref{fig.projmass}, where we plot contours of the mass within a uniform box that extends beyond 500\,kpc for all clusters. Although we plot these contours across a large region to show the general mass distribution of the cluster, the strong lensing constraints of each cluster are limited to a region that is spatially smaller than 500\,kpc from the BCG. This limitation means that the estimation of cluster mass at larger radii carries a high level of uncertainty that we cannot restrict without the introduction of other modeling techniques (i.e. weak lensing). We thus restrict our mass measurements to 200 and 500\,kpc (see \autoref{tab.clustermass} for mass estimates) and leave any inspection of larger radii to future work. 
We additionally overplot the density contours from the Chandra X-ray observations listed in \autoref{tab.xray}. These data are available for 8 out of the 14 clusters. In most cases, the center of the X-ray peak aligns well with the center of the mass peak of the cluster, except for PSZ1G091, which displays a noticeable misalignment that is discussed in \autoref{sec.PSZG091}. We show these data together to emphasize the potential of the SLICE program to make use of multi-wavelength observations when studying the physical and dynamical properties of the observed clusters.

%%%%%%%%%%%%%%%%%%%%%%%%%%%%%%%%%%%%

\subsection{New Constraints}

\begin{figure*}[hbtp!] 
   \centering 
        \includegraphics[width=0.8\linewidth]{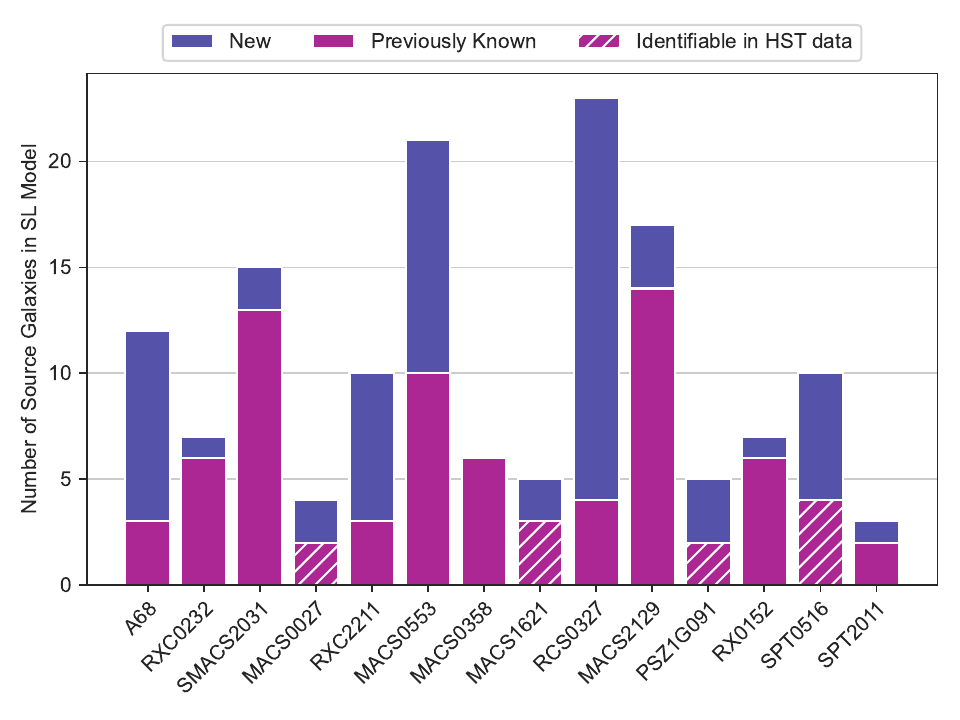}
        \caption{\label{fig.SLsystemcount} The number of \referee{sources} used to construct the SL models in this paper is broken down to demonstrate the effect of \jwst~imaging on the identification and use of constraints in the modeling process; this number does not include candidate \referee{sources}. The total number of \referee{sources} in the paper is given by the size of the whole bar for each cluster. The bar is then subdivided into the magenta section, which represents lensed \referee{source galaxies} that were previously identified in other publications for these clusters, while the purple section represents the number of new lensed \referee{source galaxies} identified in the SLICE-\jwst~images. Four clusters in this paper do not have published or public SL models available: MACS0027, MACS1621, PSZ1G091, and SPT0516. For these clusters, two of the authors independently inspected the available \hst~imaging to identify \referee{strongly-lensed galaxies} without the use of \jwst. The number of \referee{sources} identifiable in \hst~data are shown in the hashed magenta section of each bar.
        }
\end{figure*} 
%%%%%%%%%%%%%%%%%%%%%%%%%%%%%%%%%%%%
The new \jwst\ imaging facilitates the discovery of new SL features that could be used to constrain the lens models. These come in three flavors: (1) complete new \referee{systems of lensed source galaxies} that were not previously identified; (2) the discovery of new counter-images of known lensed galaxies or confirmation of suspected candidates; and (3) the identification and mapping of previously unresolved features and substructure within arcs, including space-resolution positioning of images previously only detected in MUSE data with no \hst\ counterpart. The number of newly discovered features varies from field to field. While the \jwst\ observations are uniform across the sample, the available archival data and literature range from as little as one frame of \hst\ imaging (e.g., SPT0516) and no previous lensing analysis through multi-wavelength \hst\ imaging and MUSE spectroscopy (e.g., clusters observed by treasury programs) and published lensing analysis for clusters that have previously been well studied. New\referee{ly discovered} lensed \referee{source galaxies have images that} are typically either too red or too faint, or both, to have been discovered without \jwst\ (e.g., \referee{Source} \#1 in SPT0516).
The identification of new clumps benefits from the high spatial resolution of NIRCam. Adding clumps as constraints multiplexes the number of constraints, and adds leverage over the lensing potential, as it probes the relative spatial magnification in resolved arcs. An example of the plethora of clumps in some of the most highly extended arcs is shown in \autoref{fig.arcsnapshots}. 
\autoref{tab.MultipleImages} lists the total number of \referee{sources}, multiple images, and \referee{positional} constraints (i.e., the sum of the multiple images of all the individual clumps) that were used for modeling each cluster.  
\autoref{fig.SLsystemcount} provides a visualization of the number of new and previously-known \referee{lensed source galaxies}. Where available, we determined the number of pre-\jwst\ strong lensing features from the literature or publicly available lens models on MAST. For clusters that do not have a public or published lens model, we assessed the pre-\jwst\ identification through a visual inspection of the archival data by two of the authors, at least one of whom was not involved in computing the \jwst-based model.
We note that the numbers presented in \autoref{tab.MultipleImages} and \autoref{fig.SLsystemcount} only represent the features that were used to constrain the lens models. In many of our fields, there are additional candidate \referee{sources} and images that were observed but not used as constraints (see \autoref{sec.clustermodels} for cluster-by-cluster account). 
 
New \referee{lensed source galaxies} were discovered in most cluster fields. In some, the number of \referee{sources} used to constrain lens models has more than doubled (notably, RCS 0327, MACS0553, A68, and RXC2211). In addition, SPT2011, PSZ1G091, and RCS 0327 include dozens of constraints from clumps in giant arcs (see \autoref{tab.clusterstats}). While the clusters presented in this work are not a representative sample, they span the range of SLICE targets in both redshift and mass. These number counts highlight the added value of \jwst\ imaging and validate our observing strategy for facilitating lensing analyses of the observed clusters. We expect that both the accuracy and precision of the lensing outputs would improve with number of lensing constraints \citep{johnsonsharon2016}. In future work, we will conduct a thorough investigation and comparison to expectations from simulations, by enforcing a systematic identification of arc candidates and a uniform use of the available clumps in lens models.

%%%%%%%%%%%%%%%%%%%%%%%%%%%%%%%%%%%%

\subsection{Lensing Strength}

%%%%%%%%%%%%%%%%%%%%%%%%%%%%%%%%%%%%
\begin{figure*}[htbp!] 
   \centering 
        \includegraphics[width=1\linewidth]{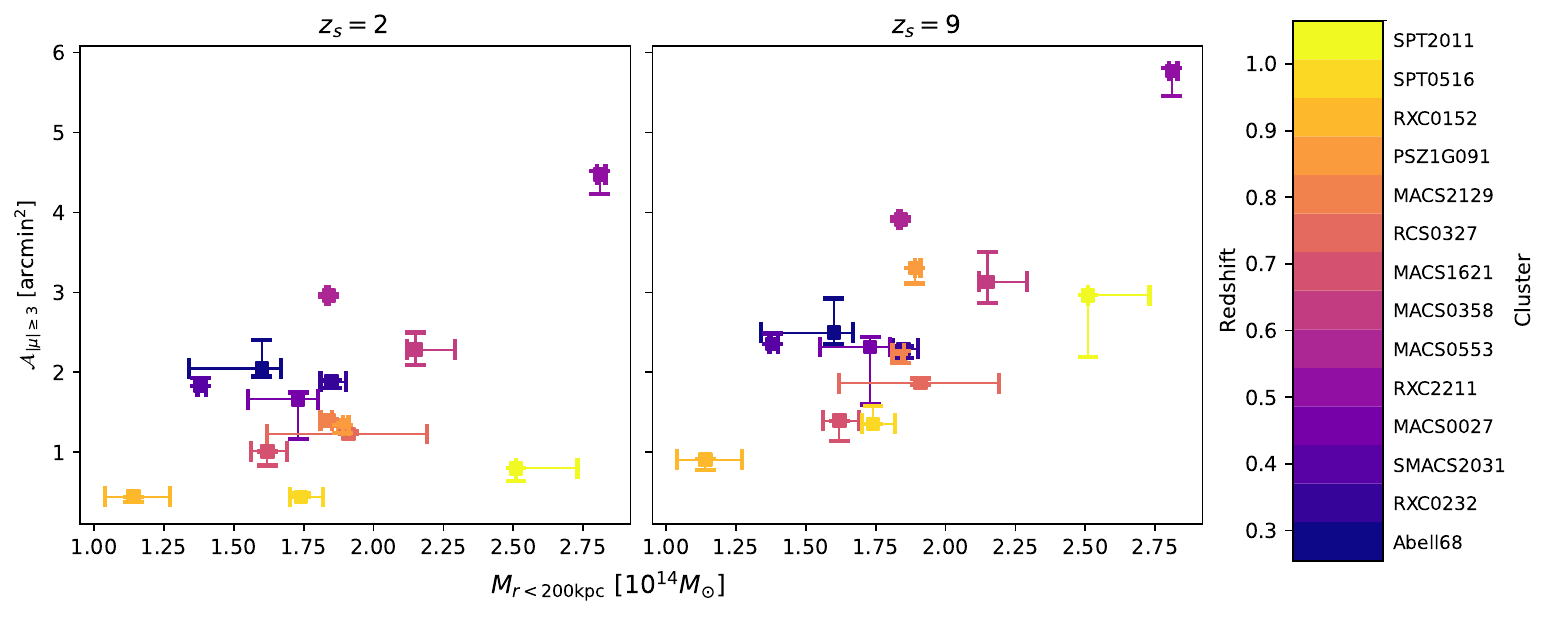}
        \caption{
        The lensing strength of each cluster, defined as the area in $\mathrm{arcmin}^2$ at which a source has an absolute magnification greater than 3, is shown for a source at $z=2$ in the left-hand plot, and $z=9$ in the right-hand plot. It is plotted relative to the projected surface mass enclosed within 200\,kpc. Each point is color-coded to correspond to a specific cluster, where the names are given on the right-hand side of the colorbar. The left-hand side of the colorbar shows the approximate redshift of the clusters, where the range of redshifts within the colorbar fully covers the upper and lower limits of the redshifts for the clusters.  
        }\label{fig.lensingstrength}
\end{figure*}

\begin{figure}[h!]
    \begin{minipage}{0.5\textwidth}
   \centering 
        \includegraphics[width=1\linewidth]{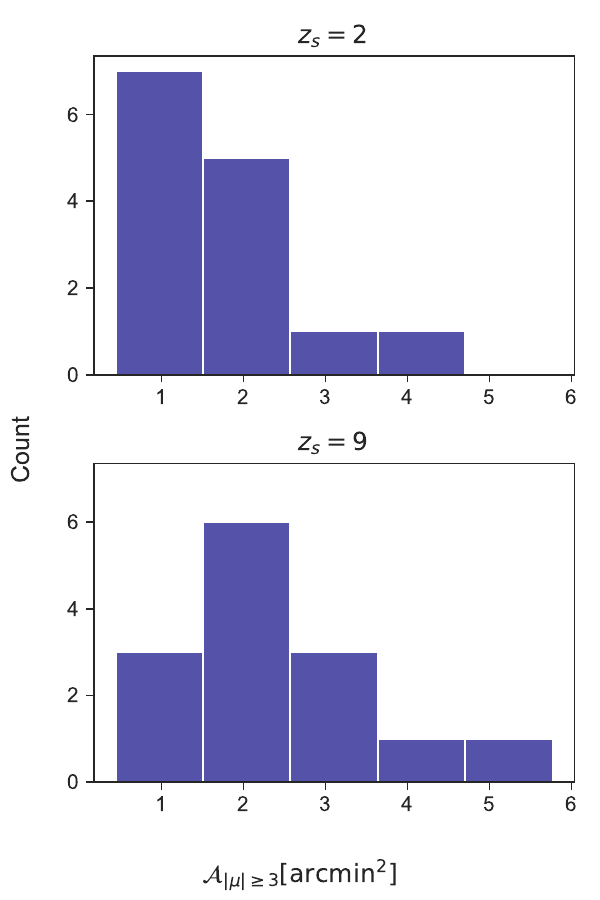}
        \caption{A histogram of the lensing strength for a source at $z=2$ (top panel) and a source at $z=9$ (bottom panel) for all clusters studied in the paper, divided into 6 equally spaced bins. The lensing strength increases for a source at $z_s=9$ relative to a source at $z_s=2$, and the distribution of lensing strength at a higher redshift is also more uniform than for a source at $z_s=2$.}\label{fig.lensinghist}
    \end{minipage}
\end{figure} 
%%%%%%%%%%%%%%%%%%%%%%%%%%%%%%%%%%%%

One of the most important applications of SL models is their use in studying the background universe.
This can be accomplished by computing the magnification of lensed sources out of the lens models.
In the era of \jwst, this is a particularly relevant quantity to be used to discover and characterize high-redshift galaxies. We examine this quantity for the 14 clusters in this paper following the method used by \cite{fox2022}, where we define the `lensing strength' $\mathcal{A}_{| \mu | \geq 3}$ of a cluster based on the size of the area in the image plane where a source at a redshift $z_s$ is magnified by a factor $\mu \geq 3$. 
This threshold is chosen as a useful break point that describes a source that has been magnified enough for identification and study while still remaining spatially close to the strong lensing constraints at the center of the cluster. Indeed, lower magnification factors further away from the cluster may not be as reliable in strong lensing models. Additionally, a source magnified by this factor can be included in studies of the high-z luminosity function (see e.g. \citealt{salmon2018, atek2023}) but remains within an acceptable range of the strong lensing area to have a reliable magnification measurement (see e.g. magnification map uncertainties in \citealt{cerny2018}). 
We choose to measure the lensing strength for a source at $z_s=2$, a common redshift for lensed galaxies used as constraints in strong lensing models, and at $z_s=9$, a common target redshift for high-$z$ galaxy searches. This also eases comparison among clusters, as it minimizes the impact of the distance between the lens and the observer. 

We compute the lensing strength of every cluster and plot this quantity relative to the mass enclosed within 200\,kpc, as shown in \autoref{fig.lensingstrength}. Generally, all clusters see an increase in lensing strength for a source at $z_s=9$ relative to a source at $z_s=2$. This is to be expected. PSZ1G091 is the most extreme example of this behavior, but we note that this jump in lensing area arises due to the addition of the cluster's Eastern mass clump at high redshift. The presence of this clump more than doubles the amplification area, as shown in the figure. However, this measurement is derived from a model without spectroscopic redshifts, and should thus be treated as preliminary until the lens model can be more precisely constrained. The sharp increase does seem to point to a potential trend, in that clusters at lower redshifts see only a minimal increase in lensing strength, while clusters at high redshifts experience a larger shift, potentially because a cluster at $z\gtrsim0.9$ is more limited in its potential to magnify sources at $z\sim2$. This trend can also be seen in the lensing strength histogram reported in \autoref{fig.lensinghist}. This demonstrates that most of the cluster lensing strength to amplify sources at $z_s=9$ is more evenly distributed, in comparison with the more concentrated area that amplifies a source at $z_s=2$.

%%%%%%%%%%%%%%%%%%%%%%%%%%%%%%%%%%%%

\subsection{High redshift detections with SLICE} \label{sec:highz}

The detection of high redshift overdensities in the background of galaxy clusters is of particular interest in clusters with strong lensing, since background galaxies can appear brighter and bigger thanks to lensing effects. However, \referee{these apparent overdensities can simply be multiple images of a strong-lensed galaxy rather than a true overdensity}. \purple{Since high redshift overdensities can potentially be an indication of proto-clusters of galaxies, careful lens modeling and spectroscopic identifications are necessary to distinguish lensed images from candidate proto-clusters } \citep[][]{zheng2014,caputi2021,laporte2022,Noirot_2023,morishita2023}.

In the field of MACS\,J0027.8+2616, a galaxy at redshift 4.45 was identified more than five times in the MUSE data cube. Of these detections, four are visible in the \jwst\ imaging, but their geometric configuration does not resemble a lensed \referee{source}. Our mass model of MACS0027 predicts that three of these detections are multiple images (shown as \referee{Source} \#30 in the model), while the other candidates do not show any signs of multiplicity.
\purple{The available information does not allow us to conclusively determine if these are \referee{multiple images of the same galaxy} or a potential proto-cluster candidate, but the detection of this high redshift overdensity is compelling grounds for future follow-up observations. Several other clusters presented in this paper show similar overdensities of relatively high redshift galaxies, which demonstrates that the large area covered by the SLICE program is well-suited to the identification of potential proto-cluster candidates. However, proto-cluster confirmation relies heavily on complementary observations (e.g., targeted spectroscopic follow-up of specific sources or additional multi-band imaging). We thus leave the broader discussion of proto-cluster discoveries from the SLICE sample to future work}. 

At intermediate redshifts, beyond the cluster lenses, typically during the so-called cosmic noon ($z$\,$\sim$\,1--3), SLICE provides exquisite spatially resolved imaging at parsec and dozens-of-parsec scales of magnified clumpy galaxies. Notably, the high sensitivity of NIRCam reveals dusty arcs that are too faint to be detected in the \hst\ WFC3/IR channel (e.g., in SPT0516)
), and that benefit from ongoing spectroscopic follow-up. Strong lensing provides the only way to access the internal physical scales that are important to characterize the individual star-forming regions at these redshifts \citep[typical clump sizes of $10-10^{2}$\,parsecs and stellar masses of $\sim 10^{5}-10^{9} M_\odot$; see][]{Claeyssens23, Rigby23}. Despite significant advancements with \jwst, the formation mechanisms and contributions of stellar clumps to galaxy mass build-up remain debated \citep[e.g.,][]{mowla2022,Claeyssens24,lahen2024,pfeffer2025}. Current and upcoming mass modeling efforts will produce the necessary underlying lens models to support the investigation of spatially resolved properties of high redshift galaxies. 

Additionally, the imaging from the SLICE program has identified a promising lensed SN candidate in SPT0516, located in a dusty galaxy that is too faint to be observed in ACS/F606W, but is revealed in NIRCam/F150W and F322W. Further investigation is needed to confirm the nature of this object, which could be a new target of choice for time-delay cosmography for precise measurements of H$_0$, independent of the distance ladder \citep[e.g.,][]{Kelly23, Pascale24}. Notably, only four lensed SNe have been confirmed to date, compared to about 300 lensed quasars \citep[e.g,][for a review]{Birrer24}.

%%%%%%%%%%%%%%%%%%%%%%%%%%%%%%%%%%%%%%%%%%%
\section{Discussion and Conclusions} \label{sec:disc}

%%%%%%%%%%%%%%%%%%%%%%%%%%%%%%%%%%%%%%%%%%%

The SLICE \jwst~program is designed to observe the co-evolution of luminous and dark matter in galaxy clusters across cosmic time, spanning a period of 8\,Gyr ($0.2<z<1.9$), for structures more massive than $M_{500} > 2 \times 10^{14}$M$_\odot$. This work features a randomly-sampled selection of 14 SL clusters from this program that span a representative range of both redshift and mass, and presents the first results for the properties of these clusters from the collaboration's SL modeling efforts. 
In nearly every cluster, the use of \jwst~imaging adds to the number of lensed sources used to build the strong lensing model (see \autoref{fig.SLsystemcount}). This increase can be as low as 1 additional \referee{lensed galaxy} and as large as 15 additional \referee{lensed galaxies}. It appears to be uncorrelated to cluster redshift or mass. \jwst\ imaging is particularly useful for revealing red, dusty galaxies, as well as very faint \referee{systems of multiple images}, both of which are essentially invisible in \hst\ imaging \citep{Perez-Gonzalez2023}.
Of the 16 models presented in this paper for the 14 clusters studied, 5 describe clusters that have not been previously published in the literature: MACS0027, MACS1621, PSZ1G09, and SPT0516. The constraints used to create these models depend on \jwst\ imaging to confirm counter-images of lensed galaxies that appear very faint in \hst\, as well as to identify red and dusty galaxies that only appear with \jwst. 

\purple{The process used to construct the lens models presented in this paper was undertaken by multiple lens modelers, who were each asked to achieve the smallest possible rms value using any visible SL \referee{constraints} in the field, as well as a satisfactory level of uncertainty given the availability of constraints (e.g. spectroscopic redshifts).} The main objective was to create a basic SL model for each cluster that could reproduce the observed \jwst\ multiple image constraints, and constrain the clusters' total mass distribution. The amount of lensed substructure used as constraints, availability of spectroscopic redshifts, and reliability of photometric redshifts varies from cluster to cluster. These models should thus be treated as preliminary results from the SLICE program, and may be updated in future work. 
We provide candidate \referee{lensed source galaxies} and substructure clumps for several of the clusters studied in this paper in \autoref{tab.MultipleImages}. These candidates are not used in this paper because they have a marginal effect on the derivation of the global mass and magnification results we discuss in \autoref{sec:results}, but studies of questions like the magnification and source-plane reconstruction of lensed galaxies, substructure statistics, and the DM distributions of these clusters may benefit from their use. We thus include them here as a reference for future work.

In general, the use of SLICE \jwst\ images improves the ability of the SL models to reproduce the observed constraints, as seen in the lower rms values reported for all models in this paper compared to previous literature. Although the difference in rms is sometimes quite small (0\farcs01 for MACS2129, for instance), the inclusion of new lensed \referee{sources} into the model (three new \referee{sources} in the aforementioned cluster) places tighter constraints on the global mass distribution, and can thus be treated as an overall improvement. In the four clusters studied in this paper where no lens models had previously been published, the specific contribution of \jwst\ is to provide identifications of \referee{lensed source galaxies} that are crucial to creating the SL model. For example, inspection of archival imaging for these clusters yielded identifications for less than half the \referee{sources} used in the final lens model presented in this paper, with the exception of MACS1621. Each of these clusters contains a red, dusty galaxy that was used as a constraint in the model, and which is crucial to resolving the shape of the overall mass distribution of the cluster.

Additionally, the ability of \jwst\ to resolve complex morphology within the multiple images aids in the construction of the lens models. For instance, inspection of \hst\ imaging for SPT2011 seems to indicate that a single lensed image is produced in the South-East (Image \#1.x.2 in \autoref{fig.allmodels}). However, examination of the SLICE \jwst\ imaging reveals that this image is actually a merging pair, with a multiplicity of up to 3 spread across the intersecting critical curve, depending on the location of the substructure clumps in the lensed galaxy. The \jwst\ imaging is also well-suited to resolving images of radial arcs that may appear within a cluster's lensing configuration. Radial arcs are a relatively uncommon strong lensing feature that lie at or near the center of a cluster's mass potential. That means they are often obscured by the light of surrounding cluster member galaxies and the BCG \citep{bartelmann2010}, making their identification reliant on spectroscopic follow-up. However, inspection of the \# Radial Arcs column of \autoref{tab.clusterstats} shows clearly that the wavelength range covered by the SLICE-\jwst\ imaging is able to either directly identify radial arcs through visual inspection, or lend compelling evidence to the presence of a radial arc- i.e. by revealing continuum emission separate from the BCG and cluster member galaxies at the location of a predicted multiple image.

Beyond the cluster lenses, the SLICE program provides exquisite spatially resolved imaging of the clumpy ISM of high-redshift lensed galaxies. This is complemented by ongoing spectroscopic follow-up, as in the case of SPT0516, which will be observed on the Magellan telescope in 2025A (U. of Michigan allocation, PI: Sharon). Combined with other facilities in the submilimeter and mid-infrared- e.g. ALMA and MIRI \citep{rigby2025, umehata2025}- SLICE has the potential to place new constraints on the physical properties governing the star formation activity in high-redshift giant molecular clouds during cosmic noon and beyond, a key epoch for galaxy evolution \citep{madaudickinson2014, weaver2023, delucia2025}.

This paper presents early results from the first clusters observed in the SLICE \jwst\ program. At the time of this paper's release, the SLICE sample covers a total of \Nobserved\ galaxy clusters, spread across a broad range of redshifts and masses. This wealth of data is designed to probe the co-evolution of luminous and dark matter across cosmic time by studying a variety of cluster properties, from dark matter and ISM distributions, to globular cluster populations, proto-cluster discoveries, and studies of the background universe. This paper thus marks the first step toward the future science that will be accomplished using the full SLICE \jwst\ sample, including providing constraints on cluster mass distributions and identifying new lensed galaxies. Since observations are still ongoing, a complete breakdown of the SLICE program, including a presentation of the whole observed cluster sample, together with statistical analyses of various cluster properties, such as lensed galaxy substructure counts, will be presented in the forthcoming survey paper (Mahler et al. 2026, in preparation). This paper will be released once observations have concluded. 

\vfill\eject
%%%%%%%%%%%%%%%%%%%%%%%%%%%%%%%%%%%%%%%%%%%
\begin{acknowledgments}

% JWST
This work is based on observations made with the NASA/ESA/CSA James Webb Space Telescope. These observations are associated with program JWST-GO-5594.
%HST
This research is based on observations made with the NASA/ESA Hubble Space Telescope. These observations are associated with programs 9290, 9770, 10493, 11591, 12100, 12166, 12267, 12313, 12362, 12371, 12884, 13412, 14096, 14098, 14630, and 15132.
% MAST
New and archival data presented in this paper were obtained from the Mikulski Archive for Space Telescopes (MAST) at the Space Telescope Science Institute, which is operated by the Association of Universities for Research in Astronomy, Inc., under NASA contract NAS5-26555 and NASA contract NAS5-03127 for JWST. The data described here may be obtained from the MAST archive at \dataset[doi:10.17909/48r6-k127]{https://doi.org/10.17909/48r6-k127}.
%Chandra
This paper employs a list of Chandra datasets, obtained by the Chandra X-ray Observatory, contained in the Chandra Data Collection (CDC) 534~\dataset[doi:10.25574/cdc.534]{https://doi.org/10.25574/cdc.534}.
% MAST
All our data products are available at MAST as a High Level Science Product via \dataset[10.17909/z77s-5c44]{\doi{10.17909/z77s-5c44}}.
% Funding
Support for program JWST-GO-5594 was provided by NASA through a grant from the Space Telescope Science Institute, which is operated by the Association of Universities for Research in Astronomy, Inc., under NASA contract NAS5-03127.
%M. Jauzac, D. Lagattuta, N. Patel, S. Werner
MJ, DL, NP, SW, JD, AE are supported by the United Kingdom Research and Innovation (UKRI) Future Leaders Fellowship `Using Cosmic Beasts to uncover the Nature of Dark Matter' (grant number MR/X006069/1).
% M. Limousin
ML acknowledges the Centre National de la Recherche Scientifique (CNRS) and the
Centre National des Etudes Spatiale (CNES) for support.
This work was performed using facilities offered by CeSAM (Centre de donnéeS Astrophysique de Marseille).
Centre de Calcul Intensif d’Aix- Marseille is acknowledged for granting access to its high performance computing resources.
% Jose M. Diego
J.M.D. acknowledges support from project PID2022-
138896NB-C51 (MCIU/AEI/MINECO/FEDER, UE) Ministerio de Ciencia,
Investigación y Universidades. 
%Mireia
MM acknowledges support from grant RYC2022-036949-I financed by the MICIU/AEI/10.13039/501100011033 and by ESF+, grant CNS2024-154592 financed by  MICIU/AEI/10.13039/501100011033, and program Unidad de Excelencia Mar\'{i}a de Maeztu CEX2020-001058-M.
%Stark
AS acknowledges support from NSF Astronomy grant 2109035.

\end{acknowledgments}

\vspace{5mm}
\facilities{JWST(NIRCam), HST(ACS, WFC3), VLT(MUSE)}

\software{astropy \citep{astropy},  
          Source Extractor \citep{sextractor},
          \lenstool\citep{jullo07},
          SAOImageDS9,
          Astrodrizzle,
          CIAO,
          Matlab,
          IDL,
          IRAF
          }

\appendix

\section{Appendix information}

%%%%%%%%%%%%%%%%%%%%%%%%%%%%%%%%%%%%%%%%%%%%%%
%CLUSTER PARAMETERS TABLE UPDATED
%%%%%%%%%%%%%%%%%%%%%%%%%%%%%%%%%%%%%%%%%%%%%%
\newpage
\startlongtable
\begin{deluxetable}{ll|lcccccccc}
\tablecaption{\purple{Lens Model Best-Fit and Median Parameters\label{tab.bestfitparams}}}
\tabletypesize{\scriptsize}
\tablehead{ 
    \colhead{Model Statistics} & \colhead{} &
    \colhead{Object} & 
    \colhead{Value} & 
    \colhead{$\Delta\alpha$ ($''$)} & 
    \colhead{$\Delta\delta$ ($''$)} & 
    \colhead{$\epsilon$} &
    \colhead{$\theta$ ($^\circ$)} & 
    \colhead{$r_{\mathrm{core}}$ ($''$)} & 
    \colhead{$r_{\mathrm{cut}}$  ($''$)} & 
    \colhead{$\sigma$ (km $s^{-1}$)}
}
\rotate
\startdata 
\para & Abell 68    & DM H1   & best & $0.46_{-0.77}^{+1.57}$ & $-2.00_{-0.81}^{+0.91}$ & $0.37_{-0.00}^{+0.34}$ & $31.4_{-1.1}^{+1.3}$ & $20.3_{-4.0}^{+0.3}$ &  [364]   & $961_{-136}^{+5}$ \\
\parb & 9.278516 &      & med. & $0.74_{-1.05}^{+1.29}$ & $-1.98_{-0.82}^{+0.90}$ & $0.54_{-0.13}^{+0.17}$ & $31.5_{-1.2}^{+1.2}$ & $19.2_{-2.9}^{+1.4}$ &  [364]   & $883_{-57}^{+84}$ \\
\parc & 9.156716 & BCG H2   & best &  [0.00]   &  [0.00]   & $0.33_{-0.00}^{+0.31}$ & $52.4_{-0.6}^{+28.3}$ & $2.5_{-1.5}^{+0.0}$ & $5_{-1}^{+9}$ & $340_{-58}^{+18}$ \\
\pard & $ 1.34 $     &      & med. &  [0.00]   &  [0.00]   & $0.52_{-0.15}^{+0.13}$ & $66.2_{-14.4}^{+14.6}$ & $1.6_{-0.6}^{+0.5}$ & $8_{-5}^{+6}$ & $312_{-31}^{+45}$ \\
\paref & $ 24, 32 $    & DM H3   & best & $-41.82_{-1.03}^{+0.61}$ & $64.04_{-0.09}^{+2.32}$ & $0.14_{-0.13}^{+0.00}$ & $150.1_{-72.0}^{+0.0}$ & $17.0_{-1.9}^{+3.7}$ & $52_{-0}^{+68}$ & $740_{-35}^{+87}$ \\
\parg & $ 0\farcs69$    &      & med. & $-42.14_{-0.71}^{+0.93}$ & $65.20_{-1.25}^{+1.16}$ & $0.03_{-0.02}^{+0.07}$ & $116.6_{-38.5}^{+27.1}$ & $17.3_{-2.1}^{+3.4}$ & $93_{-23}^{+26}$ & $760_{-56}^{+67}$ \\
    &     & Gal. H4   & best &  [27.37]   &  [22.09]   & $0.42_{-0.27}^{+0.01}$ & $23.6_{-9.1}^{+27.9}$ & $13.0_{-3.2}^{+7.6}$ & $101_{-19}^{+5}$ & $368_{-117}^{+62}$ \\
    &     &      & med. &  [27.37]   &  [22.09]   & $0.30_{-0.15}^{+0.12}$ & $30.3_{-15.7}^{+21.2}$ & $15.5_{-5.7}^{+5.0}$ & $94_{-11}^{+13}$ & $374_{-123}^{+56}$ \\
\enddata
\tablecomments{Parameters for the best-fit \purple{and median} results of all clusters modeled with \lenstool. The first two columns of the table are labeled as 'Model Statistics', and provide the following information: the cluster name; the RA and Dec of the coordinates for the BCG, given in degrees; the best-fit reduced $\chi^2$; the number of degrees of freedom and the number of free parameters ($\nu$,~N$_{\mathbf{Free}}$); and \textit{sigpos}, which represents the positional uncertainty for the multiple images that could be used to generate a model with a $\chi^2$ equal to 1. The remaining columns give the the values for the best-fit model and the median model, which are distinguished by the keyword in the 'Value' column as 'best' and 'med.' respectively. $\Delta\alpha$. and $\Delta\delta$ are defined in relation to the BCG of the clusters. Position angles are measured north of west, and the ellipticity $\epsilon$ is defined as $(a^2-b^2)/(a^2+b^2)$. Error bars correspond to the $68\%$ confidence level. Non-optimized values are indicated by brackets. Cluster names are abbreviated in this table for readability. The designations for individual potentials associated with a cluster-scale dark matter halo (DM), the BCG, or a galaxy-scale mass halo (Gal.) are abbreviated as H1, H2, etc. The properties of the reference galaxy used to scale the cluster member mass parameters are given in the row designated \lstar\ The decimal precision in each column is chosen for readability; exact values for these parameters can be retrieved from the model files hosted on the Strong Lensing Cluster Atlas Data Base. }
\end{deluxetable}

%%%%%%%%%%%%%%%%%%%%%%%%%%%%%%%%%%%%%%%%%%%%%%
%MULTIPLE IMAGES TABLE  
%%%%%%%%%%%%%%%%%%%%%%%%%%%%%%%%%%%%%%%%%%%%%%
\startlongtable
\begin{deluxetable}{lllccc} 
\tablecolumns{6} 
\tablecaption{List of multiple images} 
\tabletypesize{\small}
\tablehead{\colhead{ID} &
            \colhead{R.A.}    & 
            \colhead{Decl.}    & 
            \colhead{$z$}     & 
            \colhead{$z_{spec}$}       & 
            \colhead{Notes}       \\[-8pt]
            \colhead{} &
            \colhead{J2000}     & 
            \colhead{J2000}    & 
            \colhead{}       & 
            \colhead{Reference}       & 
            \colhead{}             }
\startdata 
  %%%%%%%%%%%%%%%%%%%%%%%%%%%%%%%%%%%%%%%%%%%
  %%%%%%%%%%%%%%%%%%%%%%%%%%%%%%%%%%%%%%%%%%%
\sidehead {{Abell\,68}}\hline\\[-5pt]
1.1.1 & 9.2808328 & 9.1578944 & 1.600  & \cite{richard2007}  &  \\
1.1.2 & 9.2803979 & 9.1566506 &  &  &  \\
1.1.3 & 9.2755277 & 9.1524037 &  &  &  \\
1.2.1 & 9.2804222 & 9.1582902 &  &  &  \\
1.2.2 & 9.2798639 & 9.1568091 &  &  &  \\
1.2.3 & 9.2747681 & 9.1528177 &  &  &  \\
\hline
2.1.1 & 9.2811551 & 9.1609147 & 1.583  & \cite{richard2007}  &  \\
\enddata 
 \tablecomments{Properties of the multiple images identified in each cluster. The image ID for each multiple image is labeled such that the first number corresponds to the \referee{lensed source}, the second number corresponds to a specific substructure clump in the image, and the third number corresponds to the image (e.g. \#\referee{Source}.\#Clump.\#Image). The RA and decl. are provided for each image in degrees, where the coordinate frame is aligned to the \jwst~WCS. The redshift and associated reference(s) for the multiple images in each cluster is also listed, where the redshift either corresponds to a spectroscopic measurement, in which case it does not include error bars, a model redshift estimate, in which case error bars are provided, or a fixed model redshift, which is denoted by brackets. Candidate lensing \referee{sources} are denoted with the letter ``c'', such that ``c1.1.1'' indicates candidate \referee{Source} \#1, Image \#1, Clump \#1. }\label{tab.MultipleImages}
\end{deluxetable}

\bibliography{bibliography}{}
\bibliographystyle{aasjournal}

%\allauthors

\end{document}